\documentclass[preprint,noshowpacs,noshowkeys]{revtex4}

\usepackage{dcolumn}
\usepackage{amsmath}
\usepackage{bm}
\usepackage{graphicx}

\usepackage{graphics}
\usepackage{longtable}
\usepackage{color}
\usepackage{epsfig}
\usepackage[normalem]{ulem}

\begin{document}
\def\Tc{$T_{\rm C}$}
\def\TN{$T_{\rm N\'eel}$}
\def\Ef{$E_{\rm F}$}
\def\BiTe{Bi$_2$Te$_3$}
\def\BiSe{Bi$_2$Se$_3$}
\def\SbTe{Sb$_2$Te$_3$}
\def\BiSbTe{(Bi, Sb)$_2$Te$_3$}
\def\SbVTe{(Sb$_{1-x}$V$_x$)$_2$Te$_3$}
\def\BiMnTe{(Bi$_{1-x}$Mn$_x$)$_2$Te$_3$}
\def\BiMnSe{(Bi$_{1-x}$Mn$_x$)$_2$Se$_3$}
\def\MnSbSbTe4{MnSb$_2$Te$_4$}
\def\MnBiBiTe4{MnBi$_2$Te$_4$}
\def\MnBiBiSe4{MnBi$_2$Se$_4$}
\def\HgMnTe{Hg$_{1-x}$Mn$_x$Te}
\def\Gbar{$\overline{\Gamma}$}
\def\MnBi{Mn$_{\rm Bi}$}
\def\muB{$\mu_{\rm B}$}
\def\dIdV{${\rm d}I/{\rm d}V$}
\def\bl{\textcolor{black}}
\def\MM{\textcolor{black}}
\def\OR{\textcolor{black}}
\def\ORa{\textcolor{black}}
\def\MMa{\textcolor{black}}

\title{Mn-rich \MnSbSbTe4: A topological insulator with magnetic gap closing at high Curie temperatures of 45--50 K}

\author{S. Wimmer$^{1,+}$, J. S\'anchez-Barriga$^{2,+}$,  {P. K\"uppers$^{3,+}$,} 
A. Ney$^1$, E. Schierle$^2$,
 F. Freyse$^{2,4}$, O. Caha$^{5}$, J. Michali\v{c}ka$^6$,  M. Liebmann$^3$, 
\bl{D. Primetzhofer$^7$,}
 M. Hoffmann$^8$, A. Ernst$^{8,9}$, M. M. Otrokov$^{10,11}$, G. Bihlmayer$^{12}$, E. Weschke$^2$, B. Lake$^2$, E. V. Chulkov$^{13,14,15,16}$, M. Morgenstern$^3$,  G. Bauer$^1$, G. Springholz$^{1*}$, O. Rader$^{2*}$}  

\affiliation{$^1$Institut f\"ur Halbleiter- und Festk\"orperphysik, Johannes Kepler Universit\"at,
			Altenberger Stra\ss e 69, 4040 Linz, Austria}
			
\affiliation{$^2$Helmholtz-Zentrum Berlin f\"ur Materialien und Energie,
			 Albert-Einstein-Stra\ss e 15, 12489 Berlin, Germany} 
						
\affiliation{$^3$II. Institute of Physics B and JARA-FIT, RWTH Aachen University, 52074 Aachen, Germany}
					
\affiliation{$^4$Institut f\"ur Physik und Astronomie, Universit\"at Potsdam, Karl-Liebknecht-Stra\ss e 24/25, 14476 Potsdam, Germany}					
							
\affiliation{$^5$Department of Condensed Matter Physics, Masaryk University, 
			Kotl\'a\v rsk\'a 267/2, 61137 Brno, Czech Republic}
			
\affiliation{$^6$Central European Institute of Technology, Brno University of Technology, Purky\v nova 123, 612 00 Brno, Czech Republic}

\affiliation{$^7$\bl{Department of Physics and Astronomy, Universitet Uppsala,
L\"agerhyddsv\"agen 1, 75120 Uppsala, Sweden}}

\affiliation{$^8$Institute for Theoretical Physics, Johannes Kepler Universit\"at,Altenberger Stra\ss e 69, 4040 Linz, Austria}

\affiliation{$^9$Max Planck Institute of Microstructure Physics, Weinberg 
2, 06120 Halle, Germany}

\affiliation{$^{10}$Centro de F\'isica de Materiales (CFM-MPC), Centro Mixto CSIC-UPV/EHU, 20018 San Sebasti\'{a}n/Donostia, Spain}

\affiliation{$^{11}$IKERBASQUE, Basque Foundation for Science, 48011 Bilbao, Spain}

\affiliation{$^{12}$Peter Gr\"unberg Institute and Institute for Advanced 
Simulation, Forschungszentrum J\"ulich and JARA,  52425 J\"ulich, Germany}

\affiliation{$^{13}$Donostia International Physics Center (DIPC), 20018 San Sebasti\'{a}n/Donostia, Spain}

\affiliation{$^{14}$Departamento de F\'{\i}sica de Materiales, Facultad de Ciencias Qu\'{\i}micas, Universidad del Pa\'{\i}s Vasco, Apdo. 1072, 20080 San Sebasti\'{a}n/Donostia, Spain}

\affiliation{$^{15}$Saint Petersburg State University, 198504, Saint Petersburg, Russia}

\affiliation{$^{16}$\bl{Tomsk State University, Tomsk, 634050, Russia}}

\date{\today}

\begin{abstract}
{\bf {Ferromagnetic topological insulators \MMa{exhibit} the quantum anomalous Hall effect \MMa{that might be used for high precision metrology and edge channel spintronics}. In conjunction with superconductors, they could host chiral Majorana zero modes which are among the contenders for the realization of topological qubits.  Recently, it was discovered that the stable 2+ state  of Mn enables the formation of intrinsic    magnetic topological insulators with A$_1$B$_2$C$_4$ stoichiometry. 
\MMa{However,} the first representative,  \MnBiBiTe4, is antiferromagnetic {with 25 K N\'eel temperature} and strongly n-doped. 
Here, we  show that  p-type  \MnSbSbTe4, previously considered topologically trivial, \MMa{is a  ferromagnetic topological insulator in the case of a few percent of Mn excess}. 
It shows (i) a \ORa{ferromagnetic hysteresis}  with \MMa{record} high Curie temperature of  45--50\,K,  
(ii) out-of-plane magnetic anisotropy and  (iii)  \MMa{a   two-dimensional Dirac cone with the Dirac point close to the Fermi level which features 
(iv) out-of-plane spin polarization} as revealed by photoelectron spectroscopy \MMa{and} (v) a magnetically induced band gap that closes at the Curie temperature as demonstrated by scanning tunneling spectroscopy.
Moreover, it displays} (vi) a critical exponent of magnetization $\beta\sim1$, indicating the vicinity of a quantum critical point. \MMa{{\it Ab initio} band structure calculations reveal that the slight excess of Mn that substitutionally replaces Sb atoms provides the ferromagnetic interlayer coupling. Remaining deviations from the ferromagnetic order, likely related to this substitution, open the inverted bulk band gap and render \MnSbSbTe4\ a robust  topological insulator and new benchmark for magnetic topological insulators.}}
\MMa{ 
} 
\end{abstract}
\maketitle
\noindent {{Corresponding authors: } 
O. Rader, email: rader@helmholtz-berlin.de, G. Springholz, email: gunther.springholz@jku.at.
} 

\noindent {$^+$ These authors contributed equally to the present work.
} 
\newpage
\vspace{2cm}

The quantum anomalous Hall effect (QAHE) offers quantized conductance and 
lossless transport without the need for an external magnetic field \cite{Onoda03}.
The idea to combine  ferromagnetism with topological insulators for this purpose \cite{CXLiu08,YuScience10,Qiao10} has fuelled
the materials science  \cite{Tokura19}. It led to the experimental discovery of the QAHE in   Cr- and V- doped \BiSbTe\   \cite{Chang13,CheckelskyNP14,KouPRL14,BestwickPRL15,Kandala2015} with  precise quantized values of the Hall resistivity \bl{down to the sub-part-per-million level} \cite{ChangCZNM15,GrauerPRB15,FoxPRB18,GoetzAPL18}. The stable 3+ configuration 
of V or Cr substitutes the isoelectronic Bi or Sb   \cite{YuScience10, YeNatComm2015, LiMPRL15} enabling ferromagnetism by coupling the magnetic moments of the transition {metal atoms}. Hence, time-reversal symmetry is broken enabling, through perpendicular magnetization, a gap opening at the Dirac point \bl{of the topological surface state} \cite{CXLiu08,YuScience10,Qiao10,Tokura19}. This gap hosts chiral edge states \bl{with precisely quantized conductivity. However, the experimental temperatures featuring the QAHE are between {$30$}\,mK \cite{Chang13,GrauerPRB15} and a few K 
  \cite{MogiAPL15,DengNatPhys2020} only,  significantly lower than} the ferromagnetic  \bl{transition temperatures} \Tc\ in these systems \MMa{\cite{Zhou2005}}.
If the temperature of the QAHE could be raised, applications such as chiral interconnects \cite{interconnects}, edge state spintronics \cite{Yasuda17,Mahoney2017} and metrological standards \cite{FoxPRB18,GoetzAPL18}  become realistic. 

\bl{One promising approach is {the so-called modulation doping in which the magnetic dopants are located  only in certain parts of the topological 
insulator. \MMa{This implies strong coupling
of the topological surface state  to the magnetic moments at a reduced disorder level}} \cite{MogiAPL15,Xiao2018}. Most elegantly, this has been realized for Mn doped  \BiTe\ and \BiSe. The} tendency of Mn to substitute 
Bi is weak, such that Mn {doping} leads to
the spontaneous formation of septuple layers with \MnBiBiTe4\ stoichiometry. These septuple layers are statistically distributed among quintuple layers of  pure \BiTe\ or \BiSe\ at low Mn concentration \cite{Rienks,HagmannNJP17} and increase in number  with \OR{increasing} Mn concentration \cite{Rienks}. \MMa{Eventually,} only septuple layers remain when the  overall stoichiometry  of \MnBiBiTe4\  or \MnBiBiSe4\  \cite{DSLee13, HagmannNJP17} \OR{is reached}. Density functional theory (DFT) calculations found that \MnBiBiTe4\ forms \bl{ferromagnetic layers with} antiferromagnetic \bl{interlayer coupling}  \cite{EremeevJAC17,OtrokovAFMTI18} \MMa{as} 
confirmed by experiments  \bl{at  low temperatures}  \cite{McQueeneyPRM19,McQueeneyPRB19,OtrokovAFMTI18,ChenPRM20}.
As a result, \MnBiBiTe4\ is an {\textit{antiferromagnetic}} topological insulator \cite{OtrokovAFMTI18,OtrokovPRL19,PRX1,PRX2,PRX3} that can exhibit axion states {\cite{Tokura19}. The QAHE, however, has only been realized in a limited way: {ultra thin flakes consisting of odd numbers of septuple layers} exhibited an anomalous Hall effect (AHE) that is nearly quantized. \MMa{This is} caused by the {uncompensated ferromagnetic septuple} 
{layer} without partner. Nevertheless, exact quantization still required {a magnetic field} \cite{DengScience20}.
A ferromagnetic AHE has also been \MMa{observed} {for  systems} with either a larger amount of quintuple layers \MMa{\cite{Vidal2019,HuSciAdv20,chen2020}} or via alloying of Sb and Bi in Mn(Bi$_{2-x}$Sb$_x$)$_2$Te$_4$
\MMa{\cite{McQueeneyPRB19,ChenPRM20,Shi2020}} or both \MM{\cite{HuarXiv2020,huan2021}}. Most notably, a nearly quantized AHE 
has been observed up to 7 K for a \MnBiBiTe4/\BiTe\ heterostructure after 
unconventional counter doping \MM{inducing vacancies by electron bombardment} \cite{DengNatPhys2020}.}

	
\bl{A central drawback of the} \BiTe\ and \BiSe\ host materials is their strong n-type doping. In contrast,  \SbTe\ is p-doped and much closer to charge neutrality \cite{Pauly2012}.  Indeed, {mixtures of \BiTe\ and \SbTe\ with} stoichiometries close to \SbTe\ have been employed for the QAHE \MMa{using} Cr and V doping \cite{Chang13,CheckelskyNP14,KouPRL14,BestwickPRL15,Kandala2015}. Magnetism of {dilute} Mn-doped \SbTe\  has initially 
been studied  by Dyck {\it et al.} obtaining $T_{\rm C}\simeq 2$\,K and perpendicular anisotropy \cite{Dyck03}.  Later, a higher $T_{\rm C}=17$\,K was  reported for 1.5\%\   Mn-doping \cite{Choi04}.  \bl{Stoichiometric bulk} \MnSbSbTe4\ \bl{provided both} antiferromagnetism \bl{(N\'eel temperature $T_{\rm N}=20$\,K)}  \cite{McQueeneyPRB19,ChenPRM20,liu2021} \bl{and ferromagnetism (\MMa{$T_{\rm C}=25-34$\,K) \cite{MurakamiPRB19,Ge2021,liu2021}} depending on the {synthesis} conditions.} \MMa{By comparison with scattering methods, it has been conjectured that this is related 
to  a remaining Mn-Sb exchange within the septuple layers \cite{liu2021,Riberolles2021,li2021,HuarXiv2020} that could even lead to spin glass behaviour \cite{li2021}}. DFT calculations, \bl{found that the perfectly ordered} \MnSbSbTe4\ \bl{is antiferromagnetic \cite{EremeevJAC17}, but} topologically trivial \MMa{\cite{ZhangDPRL19,ChenBo19,Lei20,LiuY20,Zhou2020}}, 
\MMa{ while Mn-Sb exchange can render the interlayer coupling ferromagnetic \cite{Riberolles2021,liu2021}}.

\bl{Here, epitaxial} \MnSbSbTe4\ is studied using spin- and  angle-resolved photoemission spectroscopy (ARPES),  \bl{scanning tunneling microscopy 
(STM) and spectroscopy (STS),}  \bl{magnetometry, x-ray magnetic circular 
dichroism (XMCD) and DFT. All \MMa{experimental} methods were performed as a function of temperature to pin down the intricate correlation between 
magnetism and non-trivial band topology essential for the QAHE. 
It is {revealed} that the material {unites} the favorable properties of a 
topological insulator with its Dirac point close to the Fermi level $E_{\rm F}$ {with that of a ferromagnetic hysteresis with out-of-plane anisotropy} and record-high} \Tc, twice as high as the $T_{\rm N}$  previously reported for antiferromagnetic \MnBiBiTe4\  and \MnSbSbTe4\  \cite{McQueeneyPRB19,ChenPRM20}. \bl{Moreover, temperature dependent STS finds a magnetic gap of 17 meV at $E_{\rm F}$  for 4.3 K that closes rather exactly at 
\Tc\ as expected for a ferromagnetic topological insulator.
By combining DFT, STM, Rutherford backscattering (RBS), and x-ray diffraction (XRD) it is uncovered that a partial substitution of Sb atoms by Mn  
is decisive to render \MnSbSbTe4\  both ferromagnetic \textit{and} topologically non-trivial}. 
 
\begin{figure}
\includegraphics*[scale=0.58]{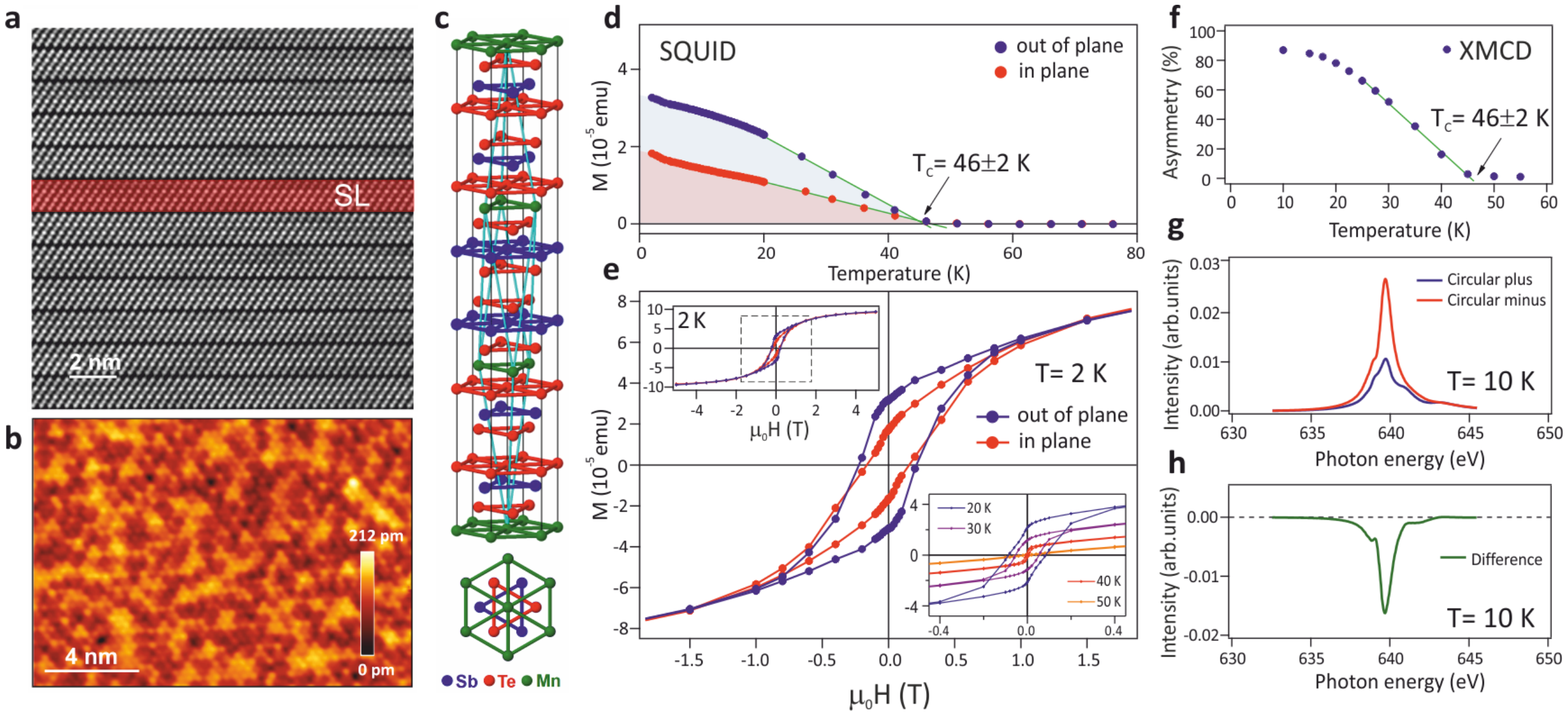}
\vspace{-5cm}
\caption{{\bf Structural \MM{and magnetic} properties of epitaxial \MnSbSbTe4.}   (a) Cross-sectional {scanning TEM} image of the septuple layer {sequence   formed} in the \MnSbSbTe4\ films \bl{on BaF$_2$ substrates}. For clarity, one of the septuples is highlighted by red color and denoted as SL. The  weak van-der-Waals-like bond between the \bl{Te-terminated} septuple layers is the natural surface termination. (b) STM image of the flat topography of this surface, recorded at $T=4.3$\,K, sample voltage of 500 mV and  current of 200 pA. \ \bl{(c) Sketch of the ideal crystal structure \MM{as side view and top view}.}\MM{(d) SQUID magnetometry as a function of temperature recorded at 10\,mT after field cooling at 5\,T. (e) Hysteresis loops probed by SQUID. The inset  shows a larger $H$ range. 
Perpendicular anisotropy can be deduced from the larger out-of-plane signal. (f) Temperature dependent XMCD signal at the Mn $L_3$-edge and the (0001) Bragg peak position recorded in x-ray scattering geometry at 0\,T after field cooling in 0.5\,T. (g) Corresponding full XMCD spectra recorded 
with opposite circular polarizations at 10\,K. (h) Difference of the spectra in (g).
Linear extrapolation in (d,f) (green lines) reveals a ferromagnetic Curie 
temperature $T_{\rm C}=46\pm2$\,K for both  experimental probes.} }
\label{Fig1a}
\end{figure}

Epitaxial \MnSbSbTe4\ films with 200 nm thickness were grown by molecular 
beam epitaxy (MBE) (Supplementary Section~\ref{sec:S1}).  Figure~\ref{Fig1a}a  shows the cross section of the \MnSbSbTe4\ lattice structure revealed by high resolution scanning transmission electron microscopy (TEM).
It consists of septuple layers (SL)  with stacking sequence Te-Sb-Te-Mn-Te-Sb-Te, \bl{Figure~\ref{Fig1a}c}. This corresponds to the \MnSbSbTe4\ stoichiometry as \bl{verified by Rutherford backscattering spectrometry (RBS), \MM{however, with a minor excess of Mn (7\,\%) compensating a minor deficiency of Sb}  (Supplement: Figure~\ref{EDFig2})}. Like with \MnBiBiTe4/\BiTe\ \cite{Otrokov2DMat}, the exchange coupling is strongly enhanced \bl{in these septuple layers} relative to a system where Mn substitutes Bi randomly. 
(24--25  K) \cite{McQueeneyPRB19,ChenPRM20}. 

The nearly exclusive formation of septuple layers in  the entire \MnSbSbTe4\ samples is confirmed by high-resolution XRD (Supplement: Figure~\ref{EDFig1}), \bl{revealing only a {minute number of residual quintuple layers}. In contrast,} TEM and XRD analysis of V-doped \BiSbTe\ shows quintuple layers only \cite{RichardsonSR17}. This highlights \MM{that septuple layers are unfavorable for V$^{3+}$.} \MMa{They} require the addition of a charge neutral transition metal$^{2+}$/Te$^{2-}$ bilayer to each quintuple layer \MMa{as} easily possible for Mn$^{2+}$ but not for V$^{3+}$.  
 Detailed XRD analysis \MMa{again} points to an exchange of Mn and Sb within the septuple layers in the 10\,\%\ range (Supplement: Figure~\ref{EDFig1}). This implies that Mn does not reside exclusively in the center of the septuples, but also to a small extent on Sb sites in the adjacent lattice planes. 
Indeed, STM images of the atomically flat and Te terminated surface of \MnSbSbTe4\ epilayers (Figure~\ref{Fig1a}b) exhibit triangular features, pointing to defects in the cation layer beneath the surface \cite{Jiang2012,Kellner2017}.  
\bl{These defects {occur} with an atomic} density of  $5-10$\,\%.  {Since 
this is significantly larger than in undoped Sb$_2$Te$_3$ films \cite{Jiang2012}, the triangles are most likely caused by  subsurface Mn atoms on Sb sites, in line with the XRD and RBS results.}  \bl{As shown by DFT below, these defects {turn out to be decisive} for the ferromagnetic interlayer coupling in \MnSbSbTe4.}    
Similar conjectures \MMa{based on scattering methods have been raised previously  \cite{liu2021,Riberolles2021,li2021,HuarXiv2020}} as well as for 
\MnBiBiTe4\ based on STM results \cite{XueSTM20}.  
%

Figure~\ref{Fig1a}d displays the temperature dependent magnetization $M(T)$ measured by a superconducting quantum interference device (SQUID). The 
measurements were recorded at 10 mT after magnetizing the sample at 2 K  by a magnetic field of 5 T perpendicular (blue) or parallel (red) to the film surface. \bl{Most strikingly, all \MnSbSbTe4\ epilayers show pronounced \textit{ferromagnetic} \MM{behavior by $M(H)$ hysteresis} loops (Figure~\ref{Fig1a}e). A record high \Tc\ of 45, 46--48 and 50 K is revealed for three independent samples (Supplement: Figure~\ref{EDFig3}), \MMa{significantly larger} as the antiferromagnetic \MMa{and ferromagnetic} transition temperatures of bulk crystals (Supplement: Table~\ref{tab1}), and twice as large as the $T_{\rm N}\le 25$ K obtained for \MnBiBiTe4\ films grown under \MM{nearly identical} conditions.} It has been crosschecked that the displayed remanent magnetization is exactly the same as in the corresponding hysteresis loops.

In particular, the \MM{large remanent magnetization observed by the bulk  
sensitive SQUID measurements excludes that it is   caused by uncompensated antiferromagnetic septuple layers only} \cite{DengScience20}. 
\MMa{The} $M(H)$ hysteresis curve (Figure~\ref{Fig1a}e), however, shows a 
rounded shape that persists up to fields much higher than those typical for domain \bl{reversals and does not saturate up to $\pm 5$\,T where the 
 magnetic moment per Mn atom is still less than 2 $\mu_{\rm B}$, similar to the result for antiferromagnetic  \MnSbSbTe4\ \cite{McQueeneyPRB19}.}
Recently, it was found that 60 T are required to fully polarize a \MnSbSbTe4\ bulk-type sample \cite{lai2021defectdriven}.
This suggests \MM{additional types of competing \bl{magnetic} orders}. 
\bl{Indeed, a kink in $M(T)$ is observed at $20-25$ K (Supplement: Figure~\ref{EDFig3}), close to the   N\'eel temperature reported earlier for \MnSbSbTe4\ \cite{McQueeneyPRB19}.}
 \bl{This implies that the high-temperature ferromagnetism is most likely 
accompanied by ferrimagnetism as {also} supported by the  relatively large in-plane hysteresis and magnetization, Figure~\ref{Fig1a}e,} \MMa{in line with observations of competing ferro- and antiferromagnetic order in bulk \MnSbSbTe4 \cite{chen2020,li2021,HuarXiv2020,liu2021,Riberolles2021,Ge2021,lai2021defectdriven}}.

\bl{The ferromagnetism is} confirmed by element specific, {\it zero field} XMCD  recorded in diffraction geometry. For \bl{these measurements}, the sample was remanently magnetized at $\sim 0.5$ T and \MMa{10 K. From spectra recorded with oppositely circularly polarized light at the (0001) Bragg peak, the} \bl{intensity difference $C_+ - C_-$ (Figure~\ref{Fig1a}h)} was deduced \bl{for 0\,T} with the photon energy tuned to the Mn-$L_3$ 
resonance \bl{(Figure~\ref{Fig1a}g)}. 
\bl{The asymmetry $(C_+ - C_-)/(C_+ + C_-)$}  directly yields the \bl{magnetization} of the \MMa{Mn}. 
Its  temperature dependence \bl{(Figure~\ref{Fig1a}f)} impressively confirms \bl{the SQUID data} with high \bl{$T_{\rm C}\simeq 46$\,K and unambiguously attributes the ferromagnetism to the Mn atoms}. 
{The record-high \Tc\ {with} magnetic easy axis perpendicular to the surface (crystallographic $c$-axis) as well as the  large coercivity} \MMa{ ($\sim 0.2$\,T),} 
($\sim10$ mT at 5 K) \cite{MurakamiPRB19} and {MnSb$_{1.88}$Bi$_{0.02}$Te$_4$} single crystals (31 mT at 2 K) \cite{ChenPRM20} Shi2020 findet auch 
0.2 T.
renders the samples very robust ferromagnets. 

\MM{Note that the spin-orbit interaction is crucial for both  out-of-plane easy axis \bl{and large coercivity} \cite{Rienks}. While it is sufficiently strong  to turn the magnetization out of plane in Mn-containing {\BiTe}, the atomic weight of Se in \MnBiBiSe4\ is too weak \cite{JSB16,Rienks}.} The present data reveals that  \SbTe\ is sufficiently heavy, i.e., spin-orbit coupling sufficiently large, to maintain the perpendicular anisotropy for high Mn content. 

\bl{To elucidate the origin of the ferromagnetism,   DFT calculations are 
employed (Supplement: Table~\ref{tab3}). They firstly highlight the differences between \MnSbSbTe4\ and \MnBiBiTe4. In both cases, the in-plane  Mn coupling within each septuple layer is ferromagnetic.} \MM{It is unlikely that the observed difference between antiferromagnetic interlayer coupling in \MnBiBiTe4\ and  ferromagnetic interlayer coupling in epitaxial \MnSbSbTe4\ 
is caused by the lowered spin-orbit interaction, since it remained large enough  for out-of-plane anisotropy}. However, the in-plane lattice constant $a$ is by $\sim2$\%\ smaller for \MnSbSbTe4. Hence, the DFT based exchange constants of \MnSbSbTe4\ {for the in-plane lattice constant $a$ determined by XRD and for $a$} expanded \bl{to the value of} \MnBiBiTe4 are compared (Supplement: Table~\ref{tab3}). \MM{While the in-plane compression increased the in-plane exchange constant $J$ between nearest neighbors 
by almost a factor of three, $T_{\rm N}$ is barely changed. The reason is 
that the enlarged in-plane \MM{overlap of Mn d states} weakens the already small, perpendicular interlayer coupling.}
Consequently, the 
energy gain of antiferromagnetism against ferromagnetism becomes as low as 0.6 meV per Mn atom. \bl{This suggests} that small structural changes \MM{along the interlayer exchange path} can induce the transition to ferromagnetic order. 

The XRD \MM{and RBS} data \MMa{imply  a Mn-Sb site exchange}. \MM{One possibility is to exchange the Mn from the central layer with Sb from the adjacent cation layers.} Here, DFT calculations demonstrate that already 2.5\,\%\ of Mn on the Sb sites, respectively, 5\,\%\ of Mn replacement by Sb in the central layer, is sufficient to swap the sign of the interlayer exchange constant (Supplement: Table~\ref{tab3}). This renders \MnSbSbTe4\ ferromagnetic at a site exchange that agrees well with both the XRD analysis and the density of subsurface defects observed by STM (Figure~\ref{Fig1a}b). A similar site exchange was recently observed \MMa{for single crystals of bulk \MnSbSbTe4\ \cite{MurakamiPRB19,liu2021,li2021}} and MnSb$_{1.88}$Bi$_{0.02}$Te$_4$  \cite{ChenPRM20}. \MM{This corroborates that the Mn-Sb site exchange is easily enabled.}
Thus, it is conjectured that  \MnSbSbTe4\  single crystals remain antiferromagnetic \cite{ChenPRM20} only for negligible Sb-Mn \MM{intermixing. However, the modelled Mn-Sb site exchange barely increases the transition temperature 
($T_{\rm N} =18$\,K $\rightarrow T_{\rm C} =25$\,K (Supplement: Table~\ref{tab3}). Instead, 5\%\  excess Mn incorporated substitutionally in the Sb layers without removing  it  from the central layer leads to $T_{\rm C}= 44$\,K (Supplement: Table~\ref{tab3}) reproducing the experimental $T_{\rm C}=45-50$\,K.
Such \Tc\ enhancement is caused by  
strengthening both the intra- and interlayer exchange constants simultaneously. The  conclusion is robust towards charge doping by up to 0.2\,\%\ Te or Sb vacancies that negligibly changes \Tc\ (Supplementary Section~\ref{sec:S11}). Note that the excess Mn in the Sb layer without Sb in the Mn layer and the resulting Sb deficiency nicely match the RBS results (Supplement: Figure~\ref{EDFig2})}
%

\begin{figure}
\includegraphics*[scale=0.4]{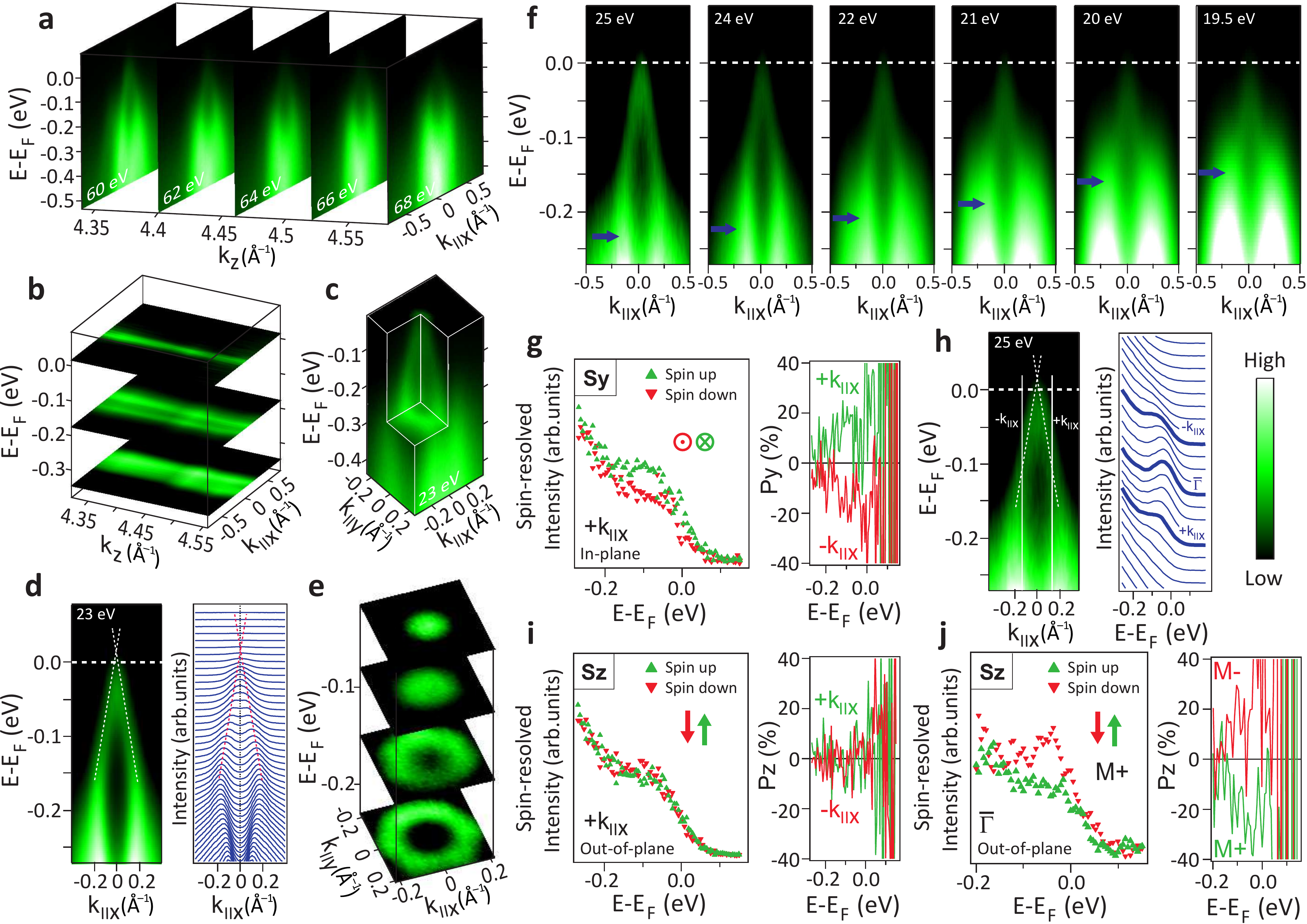}
\caption{{\bf Topological properties of \MnSbSbTe4\ revealed by ARPES.}   
\MM{(a) ARPES maps along $E(k_{\parallel x})$ for five different photon energies $h\nu$ displayed at the deduced $k_z$ values. (b) Same data as (a) displayed as ($k_{\parallel x}$, $k_z$) maps via interpolation featuring negligible dispersion along $k_z$. The $k_z$ range covers the whole Brillouin zone.
(c) Full 3D representation of the surface Dirac cone} with ${k}_{\parallel {\textrm x}}$ pointing along the $\overline{\Gamma}$-$\overline{\rm K}$ 
direction. (d) Energy-momentum dispersion of the Dirac cone (left) and corresponding momentum-distribution curves (right) \bl{recorded at $h\nu$=23\,eV}. Dashed, red lines \MM{along maxima result from fits to the data (Supplementary Section~\ref{sec:S8}). They extrapolate to} a Dirac point 20\,meV above the Fermi \MM{level}. {(e) Constant-energy cuts of the Dirac cone.}  \bl{(f) \MM{ARPES maps along $E(k_{\parallel x})$ at different photon energies} featuring a strong dispersion of a bulk valence band (blue arrows) with photon energy (i.e., $k_z$). 
(g,i) Spin-resolved ARPES at} $k_{\parallel}$ as marked in (h), hence, crossing the surface Dirac cone. Left: spectra for the two spin channels at 
one $k_{\parallel ,x}$. Right: spin polarization for both $k_{\parallel ,x}$, i.e., $\pm k_{\parallel ,x}$. (g) In-plane spin direction $S_y$ perpendicular to $k_{\parallel}$. (i) Out-of-plane spin direction $S_z$. Data 
for $S_x$ in Supplement: Figure~\ref{EDFig4}. (h) Left: surface Dirac cone with marked $k_{\parallel}$ of the spectra in (g,i) and dashed line along intensity maxima as deduced by fitting (Supplementary Section~\ref{sec:S8}). Right: energy distribution curves showing where g,i,j were measured. (j) Spin polarized ARPES recorded at ${\rm \overline{\Gamma}}$ and showcasing an out-of-plane spin polarization that reverses sign with reversal of the sample magnetization at 30 K (M$^+$ and M$^-$, right).}
\label{Fig2}
\end{figure}

Next, the topological properties of the \MM{epitaxial \MnSbSbTe4\ are probed recalling that purely stoichiometric \MnSbSbTe4\ } was predicted to be topologically trivial \cite{ZhangDPRL19,ChenBo19,Lei20,LiuY20}. It becomes a topological insulator only by replacement of more than half of the Sb by Bi  \cite{ChenBo19} or by compressing the lattice   by 3 \% \cite{ZhangDPRL19}.
\bl{However, the ARPES data from the \MnSbSbTe4\ epilayers}   \bl{ reveal 
{the existence of a surface state } with the dispersion} of a Dirac cone along the wave vector parallel to the surface ${\bf k}_\parallel$
(Figure~\ref{Fig2}a,c,d,f,h). Varying the photon energy ({Figure~\ref{Fig2}a,b}, Supplement: Figure~\ref{EDFig4a}) to tune the electron wave number perpendicular to the surface, {k}$_z$,  once through the whole bulk Brillouin zone (Supplement: Table~\ref{tab2}),  reveals no dispersion evidencing the 2D character of the Dirac cone. This is contrary to the lower-lying 3D bulk bands that strongly disperse with photon energy  (Figure~\ref{Fig2}f, arrows).  
Spin-resolved ARPES {of the 2D Dirac cone} showcases a helical in-plane spin texture \bl{ when probed away from the \Gbar\ zone center, i.e., it {exhibits the characteristic} reversal of spin orientation with the sign of ${\bf k}_\parallel$ (Figure~\ref{Fig2}g,i, Supplement: Figure~\ref{EDFig4})}. \MM{This spin chirality is a key signature} of a topological surface state. In addition, a pronounced out-of-plane spin polarization (about 
$25$\,\%)  occurs at the \Gbar\ zone center in the vicinity of $E_{\rm F}$ in the remanently magnetized sample and reverses sign when the sample is remanently magnetized in the opposite direction (Figure~\ref{Fig2}j). \MM{Such out-of-plane spin texture at \Gbar\ is evidence for a magnetic gap opening at the Dirac point \cite{Henk2012}.} 
{Combined, the ARPES results \MM{demonstrate} that \MnSbSbTe4\ is} a ferromagnetic topological insulator with \MMa{clear fingerprints of a} magnetic gap at \Gbar.  
Favorably, the Dirac point of the topological surface state is rather close to $E_{\rm F}$. Extrapolation of the observed linear bands, deduced by 
Lorentzian peak fitting  (Figure~\ref{Fig2}d,h) yields a position of the Dirac point $E_{\rm D}$ of only $20\pm7$  meV  above \Ef\ \MMa{at 300\,K} 
\MM{(Supplementary Section~\ref{sec:S8})}. 
%

\begin{figure}
\includegraphics*[scale=0.77]{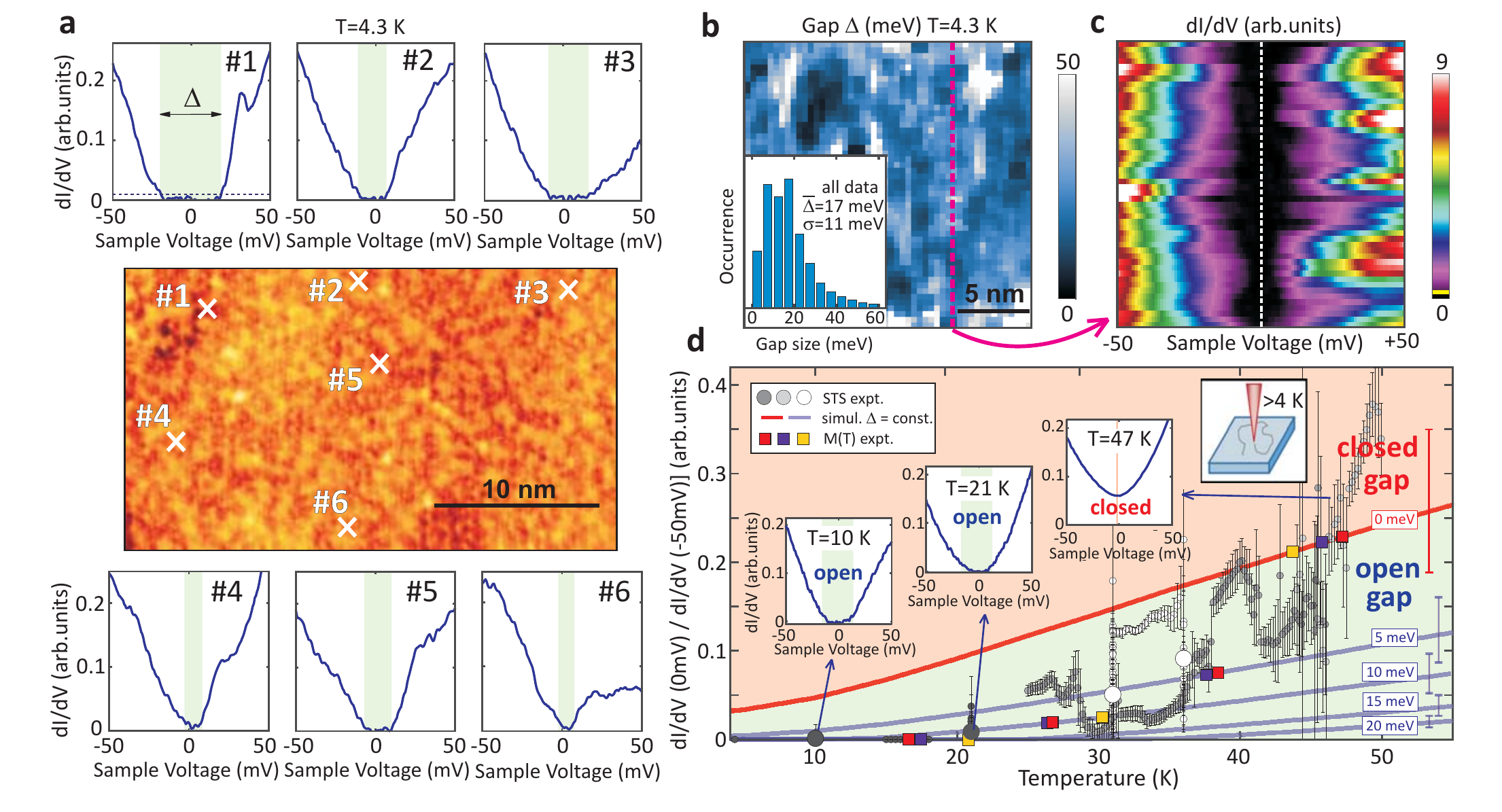}
\caption{{\bf Spatial variation of the magnetic energy gap and closing at 
\Tc.} 
{ (a) Center: STM image of \MnSbSbTe4\ after UHV transfer from the MBE, $V$ = 0.5\,V, $I$= 0.2\,nA, $T$= 4.3\,K. Surrounding: $dI/dV$ spectra recorded at the labelled crosses. Dashed line in the upper left spectrum: noise threshold used for determination of gap size $\Delta$ (Supplementary Section~\ref{sec:S9}).
 (b) Spatial map of gap size $\Delta$. Inset: 
 histogram of $\Delta$ resulting from three distinct areas, average gap size $\overline{\Delta}$ and standard deviation $\sigma$ are marked.
(c) \dIdV\   spectra (color code)
along the magenta dashed line in (b). Yellow line at the bottom of the color code bar: threshold for gap determination in (a,b). (d) Small dots: Ratio $R$=$[dI/dV(V$=$0\hspace{1mm} {\rm mV})]/[dI/dV(V$=$-50\hspace{1mm} {\rm mV})]$ deduced from $dI/dV(V)$ curves at different $T$.
Different {grey shades}: different cooling cycles. 
Each point belongs to a single position on the sample surface within 1 nm.
{Error bars: standard deviation resulting from multiple curves measured at the same position. Large grey dots: average values at the corresponding 
$T$ \MM{ representing} spatially averaged gap sizes $\overline{\Delta} (21$\,K$)$=$11\pm 5$\,meV,  $\overline{\Delta} (31$\,K$)$=$6\pm 1$\,meV, and $\overline{\Delta} (36$\,K$)$=$4\pm 1$\,meV} (Supplementary Section~\ref{sec:S9}).}
\bl{Grey lines: $T$-dependence of the ratio $R$ for the marked gap sizes $\overline{\Delta}$, deduced by convolving the measured $dI/dV(V)$ at 4.3 
K with the Fermi distribution  
 {(Supplement: Fig.~\ref{EDFig5}b).} The area above {$\Delta$=$0$ meV} exhibits such a large $R$ that the existence of a gap is excluded. Error bars on the right mark a standard deviation due to variation of the \dIdV\ curves at 4.3 K (Supplement: Fig.~\ref{EDFig5}b).
Colored squares: gap values deduced from XMCD (yellow, Fig.~\ref{Fig1a}f) 
and SQUID (red, violet, Supplement: Fig.~\ref{EDFig3}) using $\overline{\Delta}(T) \propto M(T)$ \cite{Rosenberg12}.  
  Insets: selected \dIdV\  curves for the points with blue arrows. Additional  curves: Supplement: Fig.~\ref{EDFig7}}}
\label{Fig3}
\end{figure}

Since a Dirac point above \Ef\ {is not} accessible for ARPES, STS is employed {to directly assess the ferromagnetic gap formed at $T < T_{\rm C} $}. 
\bl{Figure~\ref{Fig3}a shows a topography image (height signal in STM) of 
the \MnSbSbTe4\  epilayer together with six STS spectra recorded at 4.3\,K  at different locations of the surface.} 
\bl{All spectra consistently reveal a gap at \Ef\, varying, however, \MMa{significantly in size}. 
{Attributing} the energy region of \dIdV$\approx 0$ as gap (Supplementary 
Section~\ref{sec:S9}), a detailed map of the gap size $\Delta$ results \MMa{that covers a larger surface region (Figure~\ref{Fig3}b) \cite{LeeIPNAS2015,ChenNJP2015}. The corresponding gap {histogram  is} displayed} as inset. 
The gap size varies in the range 0--40\,meV \MM{with mean 17\,meV exhibiting a spatial correlation length of 2\,nm as observed consistently in three distinct areas (Supplement:Figure~\ref{EDFig6}).} 
A corresponding set of \dIdV\ curves recorded along the dashed line in Figure~\ref{Fig3}b is depicted in Figure~\ref{Fig3}c and showcases the small-scale band gap fluctuations. Likely, the spatial variation of $\Delta$ is caused by the spatially varying subsurface defect configuration, \MMa{i.e., by the Mn atoms} at Sb {lattice sites}. \MM{Indeed, the gap fluctuations appear on the same length scale as the topographic features in Figure~\ref{Fig1a}b.}   The average gap center position is only 0.6 meV above 
\Ef\ (Supplement: Figure~\ref{EDFig6})} with
{small} \MM{discrepancy to $E_{\rm D}-E_{\rm F}=20$\,meV as} determined 
by ARPES, which might be caused \bl{by slightly different growth conditions of the samples, \MM{by different temperatures of the two measurements (4.3\,K versus 300\,K)} or} by larger-scale  potential fluctuations {\cite{BeidenkopfNP11,Pauly2015}. 

\bl{To prove that the energy gap is of magnetic origin \cite{Rienks}, the 
temperature dependence $\Delta(T)$ is probed by STS. This has not been accomplished yet for any magnetic topological insulator {because at higher temperatures}  {$k_{\rm B}T \ge \Delta/5$ the STS gap $\Delta$} is increasingly smeared by the Fermi-Dirac distribution. This leads to a small, i.e., non-zero} tunneling current at voltages within the band gap \cite{Morgenstern03}.} A direct deconvolution of the local density of states (LDOS) and the Fermi distribution function would require an assumption on the shape of the LDOS as function of \MMa{energy. Such an} assumption is not
justified because a \MMa{significant} spatial variation of the d$I$/d$V$ curves is observed at 4.3 K (Figure~\ref{Fig3}a--c), as found consistently in other magnetic topological insulators  \cite{SessiNC16,LeeIPNAS2015}. Therefore, a {new method to derive} $\Delta$ at elevated $T$ is established. For this purpose, the ratio between \dIdV\ at $V$=$0$ mV and \dIdV\ \MM{at larger $V$, outside the region of the band gaps observed at 4.3\,K, is employed (Supplementary Section~\ref{sec:S9d})}. Figure~\ref{Fig3}d  displays the ratio $R=$ [\dIdV(0 mV)]/[\dIdV($-50$ mV)] for a {large} number of \dIdV\ spectra recorded at {temperatures varying  from 4.3\,K to 50\,K} (small dots). Selected \dIdV\ {spectra at 10, 21 and 47\,K are shown as insets  (more \dIdV\ data  in Supplement: Figure~\ref{EDFig7}).}
\bl{{One can see that up to 20\,K, \MM{the STS ratio $R$ is zero  and thus \dIdV$=0$\,nS} at zero bias, directly evidencing the persistence of the gap. At higher temperatures the $R$ values gradually increase due to the temperature broadening effect along with a decreasing magnetic gap size as the temperature approaches \Tc . To {separate}  these two effects,  the temperature dependence of the STS ratio for fixed gap sizes $\Delta(4.3 {\rm\, K})$  = 0, 5, 10, 15 and 20 meV is modelled  
by convolving the \dIdV\ curves recorded at 4.3 K  (Figure~\ref{Fig3}a--c) with the Fermi distribution, \MM{ while capturing their gap sizes $\Delta$} (Supplement:  Figure~\ref{EDFig5}). The corresponding model results are represented by solid grey lines in Figure~\ref{Fig3}d, where the red line marked with 0 meV indicates how the STS ratio $R$ evolves with temperature when the gap is zero. 
Clearly, up to $T$ = 47 K, the experimental STS ratios $R$ stay below this line. This evidences that the gap remains open up to \Tc\, whereas above {the red line} the gap is closed. It directly demonstrates the magnetic origin of the gap \cite{Rienks} due to the ferromagnetism of \MnSbSbTe4.}}


\bl{Comparing experimental data points with the calculated lines reveals that the gap size continuously decreases as the temperature approaches \Tc\ and closes rather precisely at  \Tc\ = {45--50 K} in line with the \Tc\ deduced from XMCD and SQUID. {As described above, the gap size varies 
spatially across the surface (Figure ~\ref{Fig3}a--c) due to local disorder. Accordingly, at higher temperatures the STS ratios also exhibit a considerable variation depending on where the STS spectra were recorded. For 
this reason, larger ensembles of data points have been recorded at four selected temperatures $T$ = 10, 21, 31 and 36 K within an area of 400 nm$^2$. From these, the average gap sizes are deduced to $\overline{\Delta} 
(21$\,K$)=11\pm 5$\,meV,  $\overline{\Delta} (31$\,K$)=6\pm 1$\,meV, and $\overline{\Delta} (36$\,K$)=4\pm 1$\,meV,} as represented by the large grey and \MM{white dots in Figure~\ref{Fig3}d}. Thus, the gap indeed 
gradually shrinks as the temperature approaches \Tc\ and closes above fulfilling the expectations for a ferromagnetic topological insulator \cite{Rienks}.  Note that the different grey shades of the small data points mark different cooling runs starting from an initial elevated temperature.  
Hence, the tip slowly drifts during {cooling} across the sample surface while measuring at varying temperature and, thus, explores variations of $\Delta$ by $T$ and by spatial position simultaneously. Accordingly, the visible trend of $\Delta(T)$ relies on the sufficient statistics of probed 
locations \MM{that is particularly adequate for the spatially averaged $\overline{\Delta}$} (large dots).} 

\bl{The conclusion that the gap closes at \Tc\ is corroborated by comparing the experimental gap evolution \MMa{with the measured magnetization $M(T)$ (Figure~\ref{Fig1a}d,f, Supplement: Figure~\ref{EDFig3}) using the relation} $\overline{\Delta}(T)\propto M(T)$, as found in {previous experiments \cite{Rienks} and theory} \cite{Rosenberg12}. Using $M(T)$ from SQUID and XMCD magnetization data (Figure~\ref{Fig1a}d,f, Supplement: Figure~\ref{EDFig3}) and the low temperature gap \MM{ $\overline{\Delta}(4.3\ \rm{K})=17$\,meV (Figure~\ref{Fig3}b), \MM{one} obtains $\overline{\Delta}(T) = \overline{\Delta}(4.3\ {\rm K})\cdot M(T) / M_0$ and, thus, straightforwardly $\overline{R}(T)$ for the magnetization data.} 
The results are presented as yellow, blue and red dots in Figure~\ref{Fig3}d  demonstrating nice agreement to the STS data. This further corroborates the magnetic origin of the gap. Since the center energy of the gap (Supplement: Figure~\ref{EDFig6}f) is close to the estimated Dirac point position in ARPES (Figure~\ref{Fig2}e,h), the gap is attributed to the topological surface state, consistent with the out-of-plane spin polarization 
near the Dirac point observed by spin-resolved ARPES (Figure~\ref{Fig2}j). Such a magnetic gap of a topological surface state close to  \Ef\ is highly {favorable} for probing the resulting \MM{topological conductivity 
{and its expected quantization}}.}    

\begin{figure}
\includegraphics*[scale=0.95]{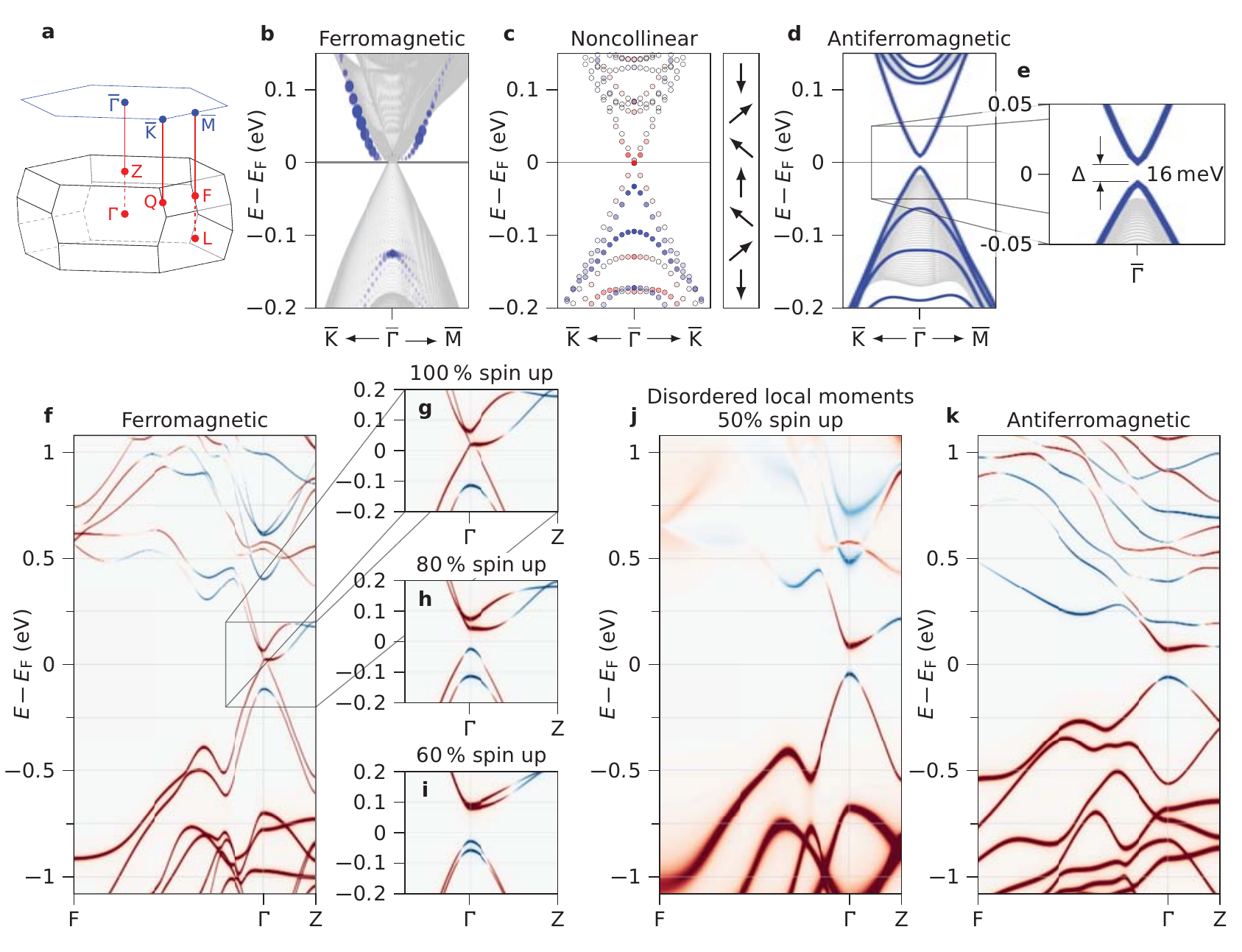}
\vspace{-1cm}
\caption{{\bf Theoretical predictions for \MnSbSbTe4.}  
(a) Brillouin zone of bulk (bottom) and surface (top) of \MnSbSbTe4\ with 
marked high symmetry points. (b--e) Band structures from DFT calculations 
of slab geometries for various magnetic configurations (marked on top). (b) Ferromagnetic interlayer coupling \MM{for a thick slab}. \MM{Blue dots: surface states identified via their strength in the top septuple layer}. (c) Non-collinear interlayer coupling (sketched on the right) for a thin slab. {Color code: out-of-plane spin polarization of the gapped Dirac 
cone. \MMa{Uncolored circles: bulk states}.} (d) Antiferromagnetic interlayer coupling \MM{for a slab (blue lines) on top of the projected bulk band structure (grey lines) with a gapped Dirac cone at \Gbar.} (e) Zoom into (d) with marked magnetic gap size of the topological surface state.
(f--k) Band structures from DFT calculation for a bulk geometry with magnetic configurations as marked on top.
Red (blue): dominating anion (cation) character as spectral function difference between anion and cation sites for each state  (Supplementary Section~\ref{sec:S11}).
(f) Ferromagnetic interlayer coupling with a topologically protected Weyl 
cone around $\rm{\Gamma}$.
(g) Zoom into (f). (h,i) same as (g) but with magnetic disorder modelled via overlap of collinear spin up and spin down states at each Mn site (Supplementary Section~\ref{sec:S11}).
(j) Same as (f), but with full magnetic disorder.
(k) Antiferromagnetic interlayer coupling. An inverted band gap appears for
(h--k) at ${\rm \Gamma}$ around \Ef\, visible by the exchanged colors of the two bands, i.e., these structures are topological insulators.}  
\label{Fig4}
\end{figure}

\bl{To  clarify the origin  of the discovered ferromagnetic topological insulator,}   the electronic  band structure of \MnSbSbTe4\ has been calculated by various DFT methods, 
considering different magnetic configurations including chemical and magnetic disorder (Supplementary Section~\ref{sec:S11}). As a general result, 
the topological insulator is \MM{reproduced by introducing} magnetic disorder. Calculating the bulk band structure, the {perfect ferromagnetic system without disorder appears as a topological Weyl semimetal} with a zero 
bulk band gap and a Weyl crossing point located  along $\rm{\Gamma}$Z about 5\%\ of the  Brillouin zone away from ${\rm\Gamma}$ (Figure~\ref{Fig4}f,g, Supplement: Figures~\ref{EDFig8}e,  \ref{EDFig9}a).  This is in agreement with recent calculations  \cite{MurakamiPRB19,Lei20}, but obviously 
disagrees with the STS \MM{and ARPES} results. On the other hand,  the  defect-free {\it anti}ferromagnetic system {\it is} found to be a topological insulator with a bulk band gap of 120 meV, Figure~\ref{Fig4}k. This is evidenced by the band inversion at $\Gamma$, indicated by the color code of the spectral function \MM{difference between cationic and anionic sites  [red (blue): dominating anion (cation) character, Supplementary Section~\ref{sec:S11}]}. This color code is also used in  Figure~\ref{Fig4}f--j. Slab calculations of the band structure (Figure~\ref{Fig4}d) indeed reveal
the topological \MM{surface state with gap at the Dirac point due to time 
reversal symmetry breaking}. \MM{ Discrepancies to earlier calculations \cite{ZhangDPRL19,ChenBo19} are discussed in Supplementary Section~\ref{sec:S11c}} 
order, {we additionally study} deviations from {defect-free, pure} ferromagnetic or antiferromagnetic order.
For more realistic modelling, more complex magnetic orders are taken into 
account deviating from  perfect ferromagnetic or antiferromagnetic order \MM{as implied by the magnetometry results} (Figure~\ref{Fig1a}e, Supplement: Figure~\ref{EDFig3}). 
The extreme case of completely disordered local magnetic moments, without 
net magnetization, showcases a bulk band gap of 135 meV with nontrivial topology as seen from the band inversion around the gap close to $\rm{\Gamma}$ (Figure~\ref{Fig4}j). Varying the degrees of magnetic disorder shows 

 that already 20\,\% of magnetic disorder breaks up the Weyl point of the 
ferromagnetic \MnSbSbTe4\ and opens an inverted band gap (Figure~\ref{Fig4}g--i, Supplement: Figure~\ref{EDFig9}). This nontrivial topology induced by magnetic disorder is robust against chemical  disorder as  Mn-Sb site exchange   that proved to be essential for inducing the ferromagnetic order in the system (Supplement: Table~\ref{tab3}). Indeed, a Mn-Sb site exchange by 5\,\% does not affect the band topology (Supplement: Figure~\ref{EDFig8}). This implies that, {contrary to recent conclusions \cite{LiuY20},}  defect engineering accomplishes simultaneously a nontrival topology and very high Curie temperature  for the \MnSbSbTe4\ system.
 
\bl{To further assess the robustness of the magnetic gap, \MM{slab calculations are employed with various magnetic disorder configurations}. The simplest case of a purely ferromagnetic slab does not lead to a Dirac cone,  since the bulk band gap vanishes (Figure~\ref{Fig4}b). The pure antiferromagnetic order, on the other hand, creates  a pronounced Dirac cone with magnetic gap of 16 meV, Figure~ 4d,e, nicely matching the average gap size found by STS (Figure~ 3b). The {magnetic} gap size turns out to be a 
rather local property caused by the exchange interaction in near-surface \MnSbSbTe4\ septuple layers. Naturally, the gapped Dirac cone forms also when the surface of an antiferromagnet is terminated by a  few ferromagnetic layers (Supplement: Figure~\ref{EDFig10}d). A relatively strong out-of-plane spin polarization at the gap edges ($\sim60$\%) is found in that case nicely matching the results of the spin-resolved ARPES data measured 
{at} 30 K {($\sim25$\%), if one takes into account the temperature dependence of the magnetization (Figure~\ref{Fig1a}d,f)}.  Moreover, for more random combinations of antiferromagnetic and ferromagnetic layers, the Dirac cone with magnetic gap persists, albeit the bulk band gap vanishes due 
to the more extended  ferromagnetic portions in that structure (Supplement: Figure~\ref{EDFig10}a).}  
Finally, for a system where the magnetic moments of adjacent septuple layers are continuously tilted with respect to each other, a gapped Dirac cone was also observed, Figure~\ref{Fig4}c. Likewise, alternate rotations of adjacent collinearly coupled Mn layers by relative angles $\ge40^\circ$ 
open a gap in the Weyl cone (Supplement: Figure~\ref{EDFig9}c--f). Hence, 
magnetic disorder turns out to be a rather universal tool to accomplish topological insulator properties for \MnSbSbTe4.

 Last but not least, \bl{it is noted that} the slope of the temperature dependent $M(T)$ shows a remarkable linear  behavior \bl{towards \Tc,} described by an effective critical exponent $\beta=0.7-1.2$. This large $\beta$ apparently persists for about half of the range between $T=0$\,K and \Tc\ {(Figure~\ref{Fig1a}d,f)}.
Such {large} $\beta$ values do not exist in any classical model ranging from $\beta\simeq 0.125$ for the 2D Ising model {to the mean-field value of 0.5}. The behavior at the classical critical point may, however, strongly change due to quantum fluctuations, which can lead to $\beta=1$ 
 in the presence of disorder \cite{Kirkpatrick} as experimentally observed 
\cite{FuchsPRB14,Sales17}.  Note that such disorder is witnessed in our samples by the spatial gap size fluctuations (Figure~\ref{Fig3}). Moreover, the magnetic phases of \MnSbSbTe4\ are indeed \MM{energetically very close  to each other according to DFT \cite{EremeevJAC17} as also indicated 
by the kink in the $M(T)$ curve (Supplement: Figure~\ref{EDFig3}, top)}.
 
In summary,  high quality epitaxial \MnSbSbTe4\ films \bl{with regularly stacked} Te-Sb-Te-Mn-Te-Sb-Te septuple layers and \MM{a small Mn excess located in the Sb layers} feature a robust nontrivial band topology at record high Curie   temperatures. DFT based band structure calculations, ARPES and STS experiments showcase a 2D Dirac cone \bl{as {nontrivial} topological surface state with a magnetic gap of $\sim 17$\,meV located very close to \Ef. The gap disappears at \Tc\ and above, signifying its magnetic origin as corroborated by an out-of-plane spin polarization at the Dirac point found by ARPES.} 
by out-of-plane spin polarization at the Dirac point found by ARPES. We {thus} conclude that the smaller spin-orbit interaction in \MnSbSbTe4\ {as} compared to \MnBiBiTe4\ is still sufficient to maintain both, the band 
inversion {and the} perpendicular magnetic anisotropy, {provided that some Mn resides in the Sb layers}.
The discovered properties are highly favorable for the quantum anomalous Hall effect and other topology-based device applications as the critical temperature is twice as large as for \MnBiBiTe4 and the Dirac point is close to $E_{\rm F}$ with small spatial variation. 

Hence, the smaller spin-orbit interaction in \MnSbSbTe4\ as compared to \MnBiBiTe4\ is still sufficient to maintain both  the band inversion and the perpendicular magnetic anisotropy \MM{in the case of a slight Mn excess}.
Ferromagnetism is triggered by a new balance of exchange interactions induced by a few percent of \MM{Mn-Sb} site exchange in the Sb layers in combination with a slight in-plane contraction as deduced by combining DFT, RBS,  STM and XRD results. The DFT calculations indicate that \MM{magnetic disorder is} essential \MM{for the magnetic topological insulator phase.}
Indeed, the magnetization features an exotic critical exponent $\beta\approx 1$ which indicates the influence of a quantum critical point, likely merging  ferromagnetic and antiferromagnetic order. 

\medskip
\textbf{Acknowledgements} \par 
We are indepted to Ondrej Man for help with lamella preparation for {TEM}. 
Financial support from the Austrian Science Funds (Projects No. P30960-N27 and I3938-N27), the Impuls-und Vernetzungsfonds der Helmholtz-Gemeinschaft under grant No. HRSF-0067 (Helmholtz-Russia Joint Research Group), the 
CzechNanoLab project LM2018110  funded by MEYS CR{, the} CEITEC Nano Research  Infrastructure, \bl{the Swedish Research Council (Project No. 821-2012-5144), the Swedish Foundation for Strategic Research (Project No. RIF14-0053),
the Spanish Ministerio de Ciencia e Innovaci\'on (Project No. PID2019-103910GB-I00), Tomsk State University (Project No. 8.1.01.2018), and Saint Petersburg State University (Project No. 51126254)} are   gratefully acknowledged. 
The work was also funded by the Deutsche Forschungsgemeinschaft within SPP1666 Topological Insulators and Germany's Excellence Strategy --- Cluster of Excellence Matter and Light for Quantum Computing (ML4Q) EXC 2004/1 --- 390534769,   and the project Mo 858/13-2 as well as by the Graphene Flagship Core 3. 

 \newpage

\section*{Supplementary Sections}
The following sections describe the experimental and theoretical methods as well as additional data supporting the conclusions from the main text.

\newpage
\section{Sample Growth}
\label{sec:S1}
\MnSbSbTe4\ films were grown by molecular beam epitaxy (MBE) on BaF$_2$(111) substrates using a Varian GEN II system. Compound \SbTe\  and  elemental Mn and Te sources were employed for control of stoichiometry and composition. Typical sample thicknesses were 200 nm. Deposition was carried out at a sample temperature of 290$^\circ$C at which perfect 2D growth is sustained independently of the Mn concentration, as verified by {\it in situ} reflection high energy electron diffraction and atomic force microscopy. The flux rates were calibrated by quartz microbalance measurements. For angle-resolved photoelectron spectroscopy (ARPES) performed at BESSY II \bl{and for scanning tunneling microscopy (STM) performed at RWTH Aachen University}, samples were transferred from the MBE system by an ultrahigh vacuum (UHV) suitcase \bl{that allowed transportation} without breaking UHV conditions. Samples used for {magnetometry and} x-ray magnetic circular dichroism (XMCD) were capped {\it in situ} after growth with amorphous Se and Te capping layers to protect the surface against oxidation. 

\section{Electron Microscopy} 
\label{sec:S2}
High-resolution scanning transmission electron microscopy ({TEM}) was performed with a FEI Titan 60-300 
Themis instrument equipped with a Cs image corrector. The {TEM} data were 
recorded with a high-angle annular dark field (HAADF)  detector and the images processed using a fast Fourier transform and Fourier
mask filtering technique for noise minimization.  Thin cross-sectional lamellae from \MnSbSbTe4\ films were prepared by focused ion beam milling (FEI Helios NanoLab 660). 

\section{Rutherford Backscattering}
\label{sec:S3}
\begin{figure}
\includegraphics*[scale=0.6]{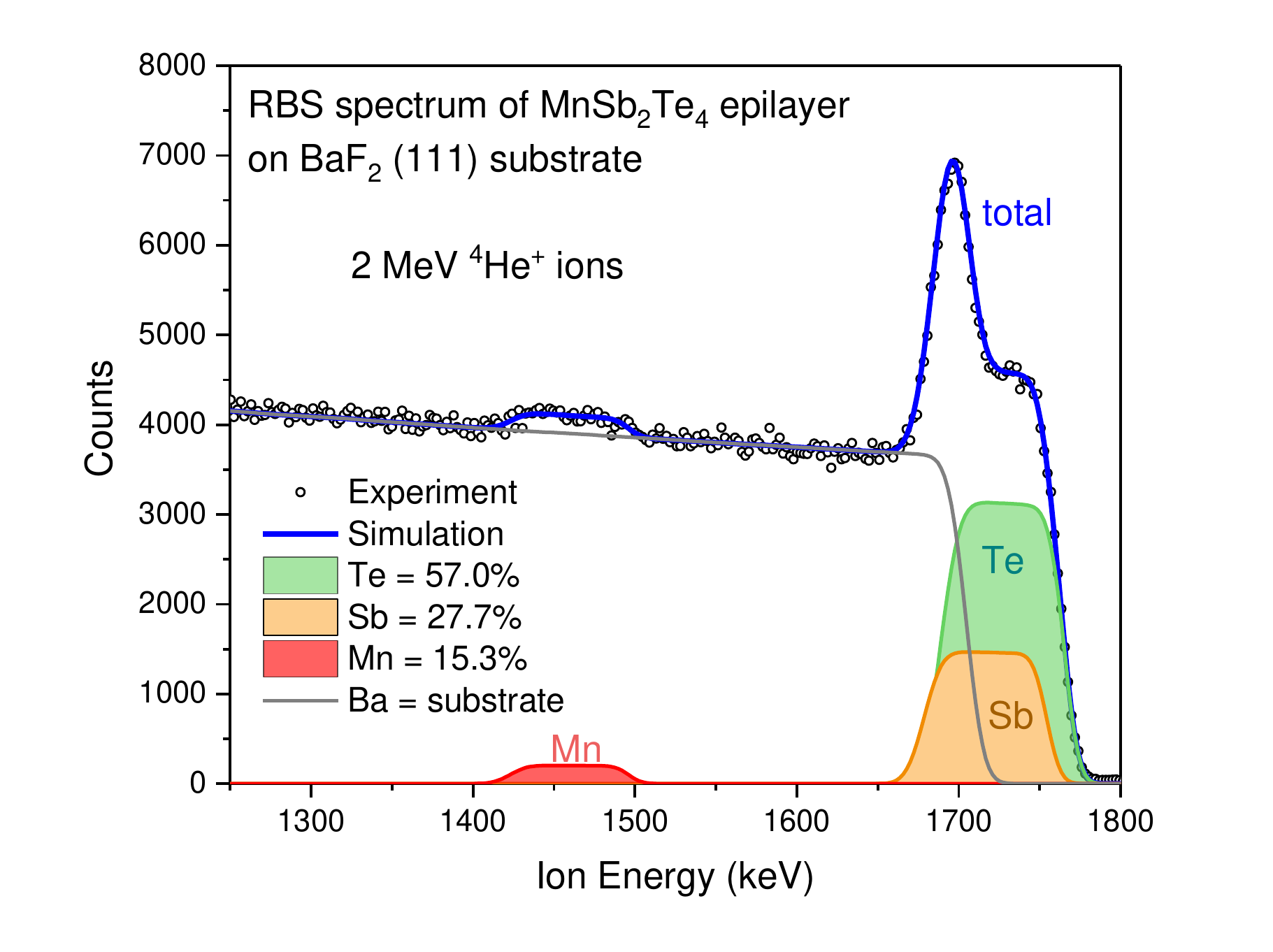}
\vspace{-0.6cm}
\caption{{\bl{\bf Chemical composition and depth profile.}}
Rutherford backscattering spectrum (open circles) measured in random geometry. The simulation of the total backscattering signal using SIMNRA (blue line) is in excellent agreement with the experimental data. The corresponding contributions of the film constituents (Te, Sb and Mn) are marked and indicated by the green, orange and red lines surrounding correspondingly shaded areas. The data reveal a slight Mn excess of $\sim 7$\,\%\ and 
an Sb deficiency of $\sim3$\,\%\ with respect to the ideal \MnSbSbTe4 stoichiometry suggesting a partial Mn replacement of Sb atoms in the Sb layer.}
\label{EDFig2}
\end{figure}

\bl{The chemical composition of the samples was determined by Rutherford backscattering spectrometry (RBS) employing a primary beam of 2 MeV $^4$He$^+$ ions provided by the 5 MV 15SDH-2 Pelletron accelerator at Uppsala University.  A solid-state detector was placed in backscattering geometry 
under an angle of 170$^\circ$ with respect to the primary beam. The beam incidence angle was randomized in a 3$^\circ$ angular interval around an equilibrium angle of 18$^\circ$ with respect to the surface normal to counteract potential channeling effects. The resulting spectra (Fig.~\ref{EDFig2}) were fitted by the SIMNRA software package \cite{Mayer14} to obtain the film composition and {confirm} its uniformity over the film cross section.} 
Analysis of the RBS data as shown in Fig.~\ref{EDFig2} yields an excellent fit (blue line in Fig.~\ref{EDFig2}) for atom concentrations of 57.0\%\ 
Te, 27.7\%\ Sb and 15.3\%\ Mn with a high uniformity throughout the film cross section. The composition is close to the nominal \MnSbSbTe4\ stoichiometry, but features a slight Mn excess of $\sim7$\%\ and a slight Sb deficiency of $\sim3$\%\ indicating the substitution of Sb atoms in the Sb layers by additional Mn atoms as corroborated by the x-ray diffraction data shown in section~\ref{sec:S4}.

\section{ X-ray Diffraction}
\label{sec:S4}

\begin{figure}
\includegraphics*[scale=0.75]{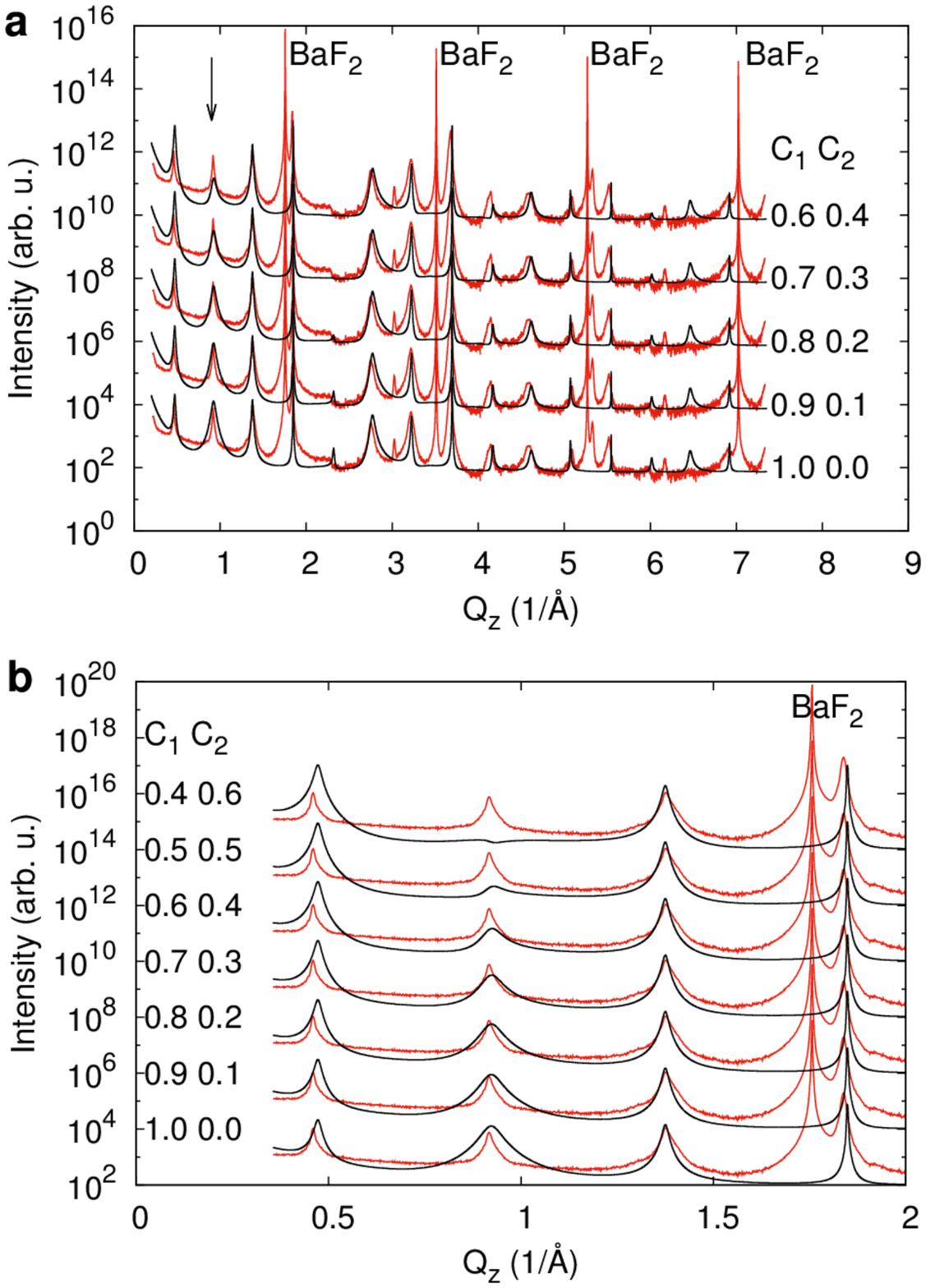}
\vspace{-4.5cm}
\caption{{\bf X-ray diffraction spectra with fits.}
(a) Experimental x-ray diffraction spectrum  of the \MnSbSbTe4\ epilayer (\bl{red} lines) compared with simulations (\bl{black} lines) using a one-dimensional paracrystal model \cite{Steiner14,Rienks} while assuming different Mn occupation in the central layer  ($C_{1}$) and  in the adjacent 
cation layers ($C_2$) of the septuple layer.  (b) Zoom into (a) showcasing the strong sensitivity of the peak at $Q_z\approx 0.9$ \AA$^{-1}$ (black arrow in (a)) to the Mn-Sb site exchange. Peaks of the BaF$_2$ substrate are marked.}
\label{EDFig1}
\end{figure}

The crystalline structure was determined by x-ray diffraction (XRD) scans 
and reciprocal space maps recorded in the vicinity of the (10$\overline{1}$.20) reciprocal lattice point. The measurements were performed using a Rigaku SmartLab diffractometer with Cu x-ray tube and channel-cut Ge(220) 
monochromator. The results are shown as red lines in Fig.~\ref{EDFig1}. The symmetric scans along the [000.1] reciprocal space direction ($c$-axis) were fitted with a modified one dimensional paracrystal model \cite{Steiner14,Rienks} (black lines in Fig.~\ref{EDFig1}).
We have  fitted the data of two samples and obtained consistent results of $a= 4.23$ \AA\ and $c=40.98$ \AA. 

We modelled the Mn-Sb site exchange by different Mn occupancies $C_{1}$ and $C_2$  within the central cation lattice planes, usually assumed to be 
occupied by Mn only, and within the outer cation lattice planes, usually assumed to be occupied by Sb only, respectively. 
For that purpose, the structure factor of \MnSbSbTe4\ reads  
\begin{equation}
F(Q_z)=\sum_n  C_n f_n(Q_z) \exp(-i  Q_z z_n),
\end{equation}
where $f_n(Q_z)$ is the atomic form factor of the $n$-th atom in the unit 
cell, $z_n$ its coordinate, $C_n$ its occupancy, $Q_z$ its position in reciprocal space and the summation runs over all atoms in the septuple layer.  For Sb, Te, and Mn the atomic form factors at $Q_z=0.9$\,\AA$^{-1}$ 
are 48.9, 49.9 and 23.6, respectively. 

The fit result that assumes the ideal septuple layer structure with regular layer sequence Te-Sb-Te-Mn-Te-Sb-Te  ($C_1$ = 1, $C_2$ = 0) is shown on the bottom of Fig.~\ref{EDFig1}a, while $C_1$ = 0.6 and $C_2$ = 
0.4, corresponding to  60\% of the Mn atoms in the central septuple layer 
and 40\% in the outer two cation layers, is shown on top.
Figure~\ref{EDFig1}b highlights the low $Q_z$ region featuring the strong 
sensitivity of the peak at $Q_z=0.9$\,\AA$^{-1}$ to the exchange of Mn and Sb. 
For this peak, the two different cation layers contribute to the structure factor with almost exactly opposite phase, while the contribution of the Te layers almost cancels. Because of the different Mn and Sb atomic form factors, the  intensity of this peak rapidly diminishes with increasing 
Mn-Sb exchange and completely vanishes for  $C_1 = C_2$ = 0.5. The best fit to the experiment is found for $C_1=0.9\pm0.1$ and $C_2=0.1\pm0.1$ evidencing a non-ideal distribution of Mn and Sb in the respective layers. 

\section{ Magnetometry by SQUID}
\label{sec:S5}
The magnetic properties were determined by a superconducting quantum interference device (SQUID) magnetometer (Quantum Design MPMS-XL) as a function of temperature from 2\,K to 300\,K. The external field {\bf H} was applied either parallel (in plane) or perpendicular (out of plane) to the epilayer surface, that itself is oriented perpendicular to the crystallographic $c$ axis. The diamagnetic contribution of the substrate was determined from the slope of the magnetization $M(H)$ \bl{recorded at high magnetic fields and 300\,K well above the Curie temperature \Tc\ of \MnSbSbTe4, 
where the magnetic contribution of the thin film is completely superseded 
by that of the thousand times thicker substrate.  The derived substrate contribution was then} subtracted from the raw data recorded at lower $T$. 
Identical sample pieces were used for in-plane and out-of-plane measurements.

\begin{figure}
\includegraphics*[scale=1.2]{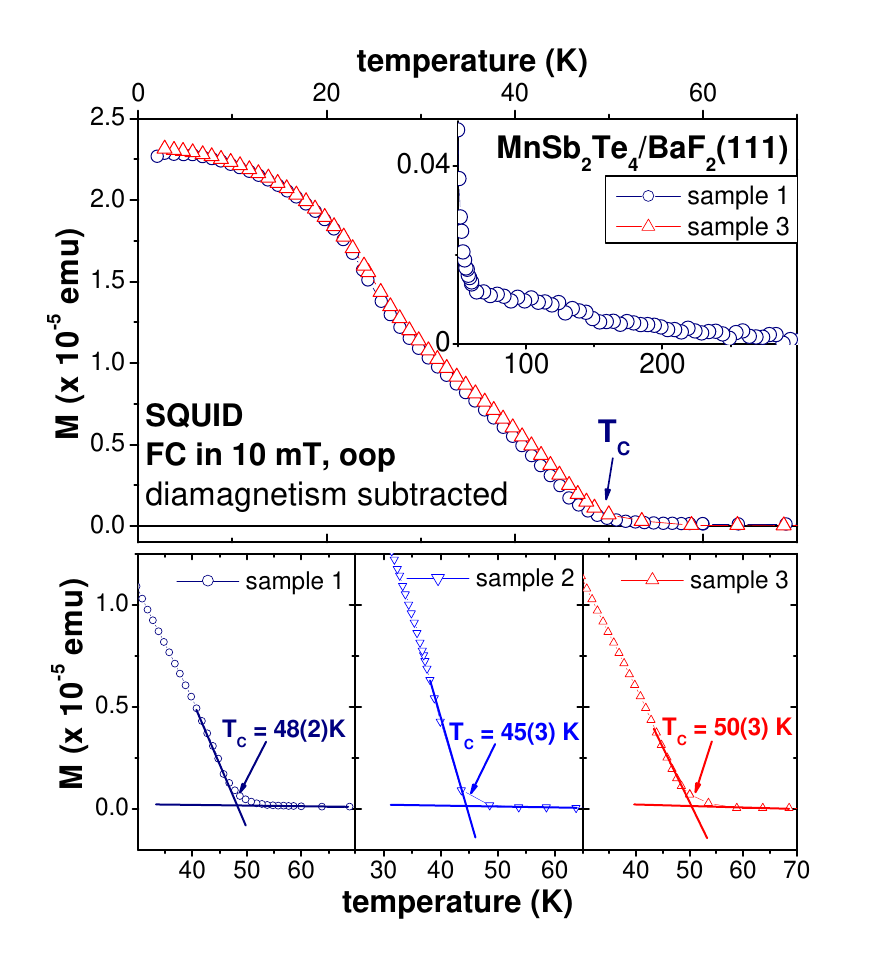}
\vspace{-0.5cm}
\caption{\bl{{\bf Magnetometry by SQUID.}}
\bl{Magnetization $M(T)$ of  three {individual \MnSbSbTe4\  epilayers measured by SQUID. $M(T)$ was measured in out-of-plane (oop) direction while 
field cooling (FC) in a field of 10 mT. 
Linear extrapolation to $M=0$\,emu was used to estimate \Tc\ as marked.
Notice the small kink in $M(T)$ around 20\,K in the top panel which indicates deviations from pure ferromagnetic order. Inset: Full  temperature dependence of the magnetization $M(T)$ up to 300 K on a vertical scale enlarged by a factor of 40. From the SQUID raw data, the diamagnetic contribution from the BaF$_2$ substrate of 0.0205 $\mu$emu/Oe was substracted.}}}
\label{EDFig3}
\end{figure}

Figure~\ref{EDFig3} shows out-of-plane $M(T)$ data while field cooling at 
10\,mT for three different samples. They reveal a quite similar \Tc\ and consistently display a ferromagnetic behavior via its remanence. Data for 
sample 1 are also shown in Fig.~\ref{Fig1a} (main text) where they give a 
slightly smaller \Tc\ within the error bars. 

Figure~\ref{EDFig3} (top) features a small kink at 20--25\,K that indicates deviations from pure ferromagnetic order possibly indicating a second phase transition that might contribute to the large critical exponent $\beta\simeq 1$ as discussed in the main text and attributed there to the vicinity to a quantum critical point.

\section{Comparison with Magnetic Properties  of Other \MnSbSbTe4 Samples}
\label{sec:S6} 
\begin{table}
\vspace{-2cm}
\includegraphics*[scale=0.9]{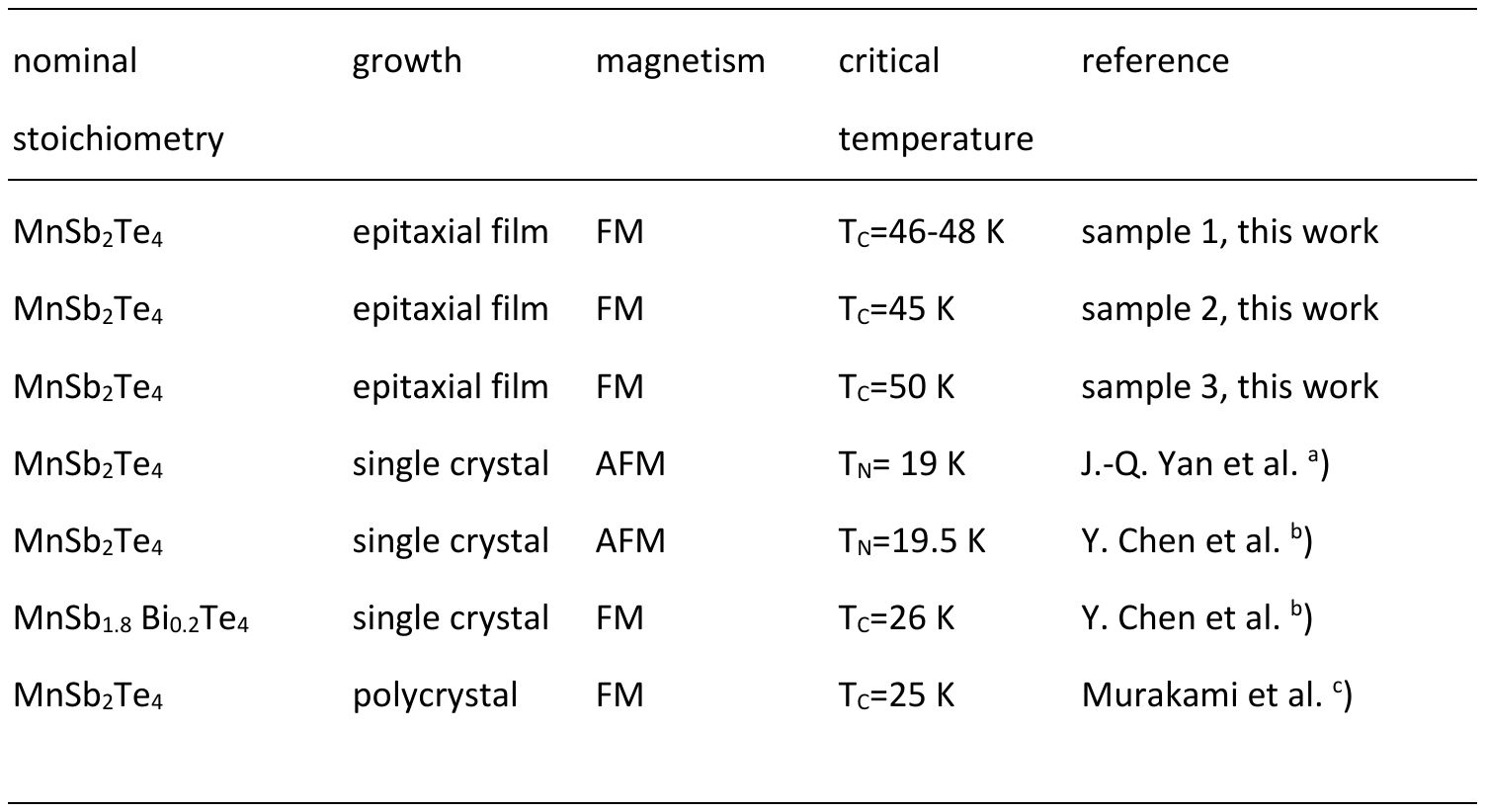}
\vspace{-18cm}
\caption{\bl{{\bf Magnetic properties of \MnSbSbTe4} for epitaxial films, 
 bulk crystals and polycrystals, comparing the ferromagnetic (FM) Curie temperatures \Tc\ determined in the present work to \Tc\ and antiferromagnetic (AFM) N\'eel temperatures $T_{\rm N}$ from the literature. $^{\rm a}$) J.-Q. Yan et al., Phys. Rev. B 100, 104409 (2019); $^{\rm b}$) Y. Chen 
et al., Phys. Rev. Mat. 4, 064411 (2020); $^{\rm c}$) T. Murakami et al., 
Phys. Rev. B 100, 195103 (2019).}}
\label{tab1}
\end{table}

  Table~\ref{tab1} compares the results of our three \MnSbSbTe4\ samples with data from the literature. The comparison highlights the exceptionally large Curie temperature of the Mn-rich epitaxial films. 

\newpage
\section{Resonant Scattering and X-ray Circular Dichroism}
\label{sec:S7}
Resonant scattering and XMCD were measured at the extreme ultraviolet (XUV) diffractometer of the UE46-PGM1 undulator beam line of BESSY II at Helmholtz-Zentrum Berlin. The XMCD signal was obtained by measuring the difference of the (0001) Bragg peak intensities for incident photons with opposite circular polarization and the photon energy tuned to the Mn-$L_3$ resonance. In this setup, the sample was field cooled down to 10\,K in an external field of about 0.5\,T provided by a removable permanent magnet. Subsequent polarization dependent and wave-vector dependent measurements were performed in zero field at various temperatures. The recorded XMCD data are thus a direct measurement of the remanent ferromagnetic polarization of the Mn moments in \MnSbSbTe4.

\section{ARPES and spin-resolved ARPES}
\label{sec:S8}
ARPES measurements were performed at 30\,K with a Scienta R4000 hemispherical analyzer at the RGBL-2 end station of the U125/2 undulator beamline at BESSY II. Light is incident under an azimuthal angle of 45$^\circ$ and 
a polar angle of 90$^\circ$ with respect to the sample surface. The light 
polarization is  linear and horizontal. Photon energies between 19 and 70\,eV were employed. Spin-resolved ARPES spectra were acquired with a Mott-type spin polarimeter operated at 25\,kV, capable of detecting both in-plane and out-of-plane spin components. Samples were transported to the RGBL-2 setup in a UHV suitcase to always  maintain pressures below $1\cdot10^{-10}$\,mbar.  Overall resolution of ARPES measurements was 10\,meV (energy) and $0.3\,^{\circ}$ (angular). Resolutions for spin-resolved ARPES were 45\,meV (energy) and $0.75\,^{\circ}$ (angular).

For the measurement in Fig.~\ref{Fig2}j, main text, the sample was magnetized in situ at 30 K by applying a pulsed magnetic field from a removable 
coil of  $\sim\pm 0.5$ T in the direction perpendicular to the surface. As usual, the spin-resolved ARPES experiment is conducted in remanence. The movement between the ARPES chamber and the preparation chamber, where the coil is situated, is a vertical movement of the cryostat so that the sample is always at 30 K.  

\begin{figure}
\includegraphics*[scale=0.6]{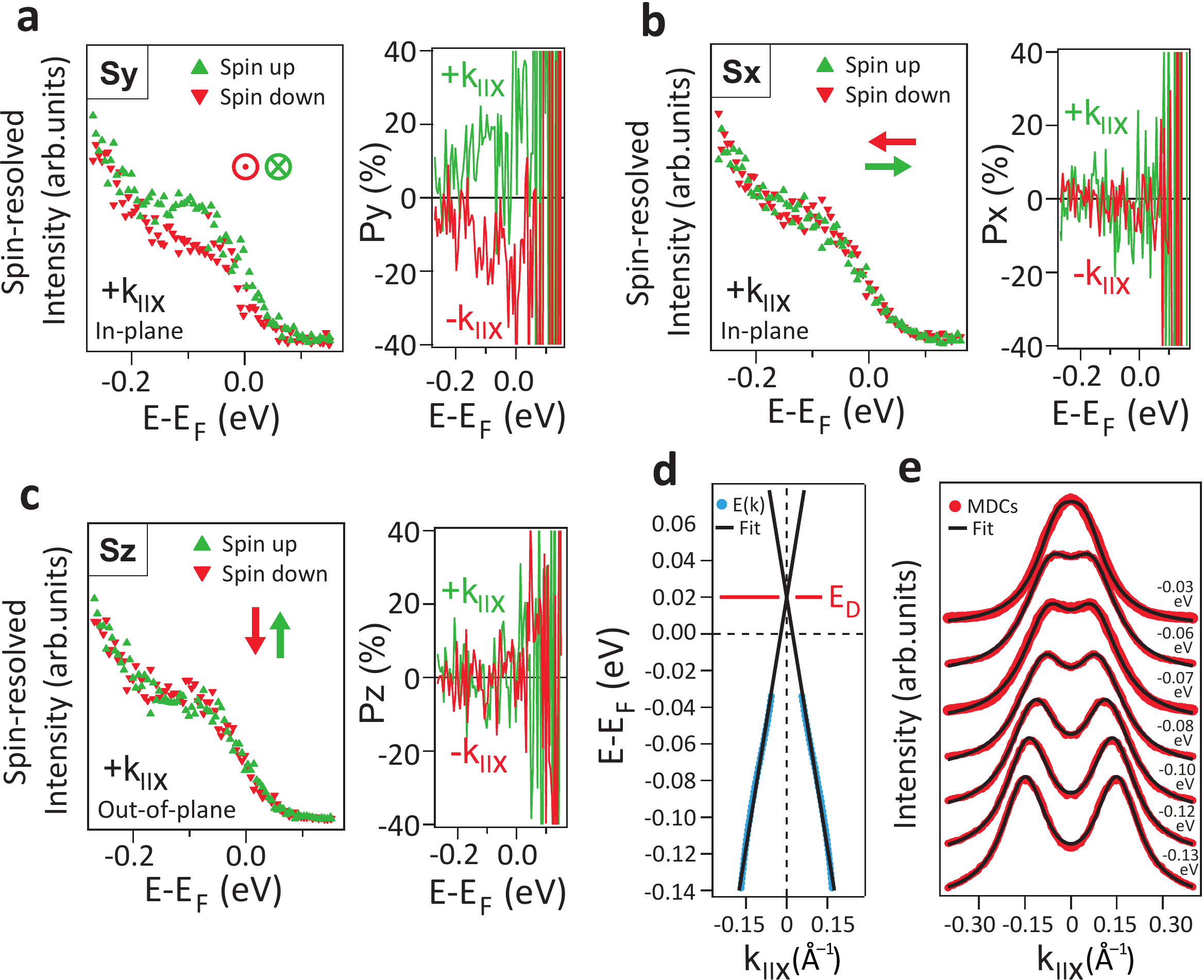}
\vspace{-0.0cm}
\caption{{\bf Spin-resolved ARPES data and fit of Dirac cone.} 
All three spin polarization components are shown for $k_{\parallel,x} \simeq \pm 0.15$\,\AA$^{-1}$ (indicated in Fig.~\ref{Fig2}h, main text). Left hand side of each subfigure shows the ARPES data for the two opposite spin channels at $k_{\parallel,x} \simeq +0.15$\,\AA$^{-1}$, while the right hand side shows the resulting spin polarization $P_i$ for both $k_{\parallel,x} \simeq \pm 0.15$\,\AA$^{-1}$. (a) In-plane spin components perpendicular to $k_{\parallel,x}$ (same as Fig.~\ref{Fig2}g, main text). (b) 
In-plane spin components parallel/antiparallel to $k_{\parallel,x}$. (c) Out-of-plane spin components. The symbols for spin direction and orientation are relative to panel (d). (d) The Dirac energy $E_{\rm D}$ is determined by the crossing of two lines given by fits to experimental data points which, in turn, were determined by Lorentzians fitted to momentum distribution curves (MDC). These fits are shown in panel (e). $h\nu=25$ eV.} 
\label{EDFig4}
\end{figure}

Figure~\ref{EDFig4} shows a complete set of spin-polarized ARPES data featuring all three orthogonal spin directions for a finite $k_{\parallel}=\pm 0.15$\,\AA$^{-1}$. Only in-plane spin polarization perpendicular to the electron wave vector $k_{\parallel}$ appears and reverses sign with a sign  change
of $k_{\parallel}$. This evidences a spin texture that rotates anti-clockwise (right-handed) around the measured Dirac cone (Fig.~\ref{Fig2}g,i, main text) with the electron spin locked to the electron momentum. Such helical in-plane spin texture is a key signature of the topological character of the Dirac cone. Note that the ferromagnetism of \MnSbSbTe4\ exhibits out-of-plane anisotropy such that bulk states cannot have in-plane spin 
polarization, which fully supports the assignment to the Dirac cone surface state. To deduce the Dirac point energy $E_{\rm D}$, the peaks within momentum distribution curves are fitted by two Lorentzians (Fig.~\ref{EDFig4}e) and the energy dependent peak maxima are connected via two fitted lines (Fig.~\ref{EDFig4}d and Fig.~\ref{Fig2}d,h, main text) that cross at $E_{\rm D}-E_{\rm F}=20\pm 7$\,meV. The binding energy range of the fit was determined by the possibility to fit separate Lorentzians, see Fig.~\ref{EDFig4}d,e. Since we do not observe any  surface band bending by 
residual gas adsorption (prominent for the n-doped systems such as  \BiSe\ and \BiTe), the Dirac point can be regarded as a rather good measure of 
the chemical potential.

\begin{figure}
\includegraphics*[scale=0.6]{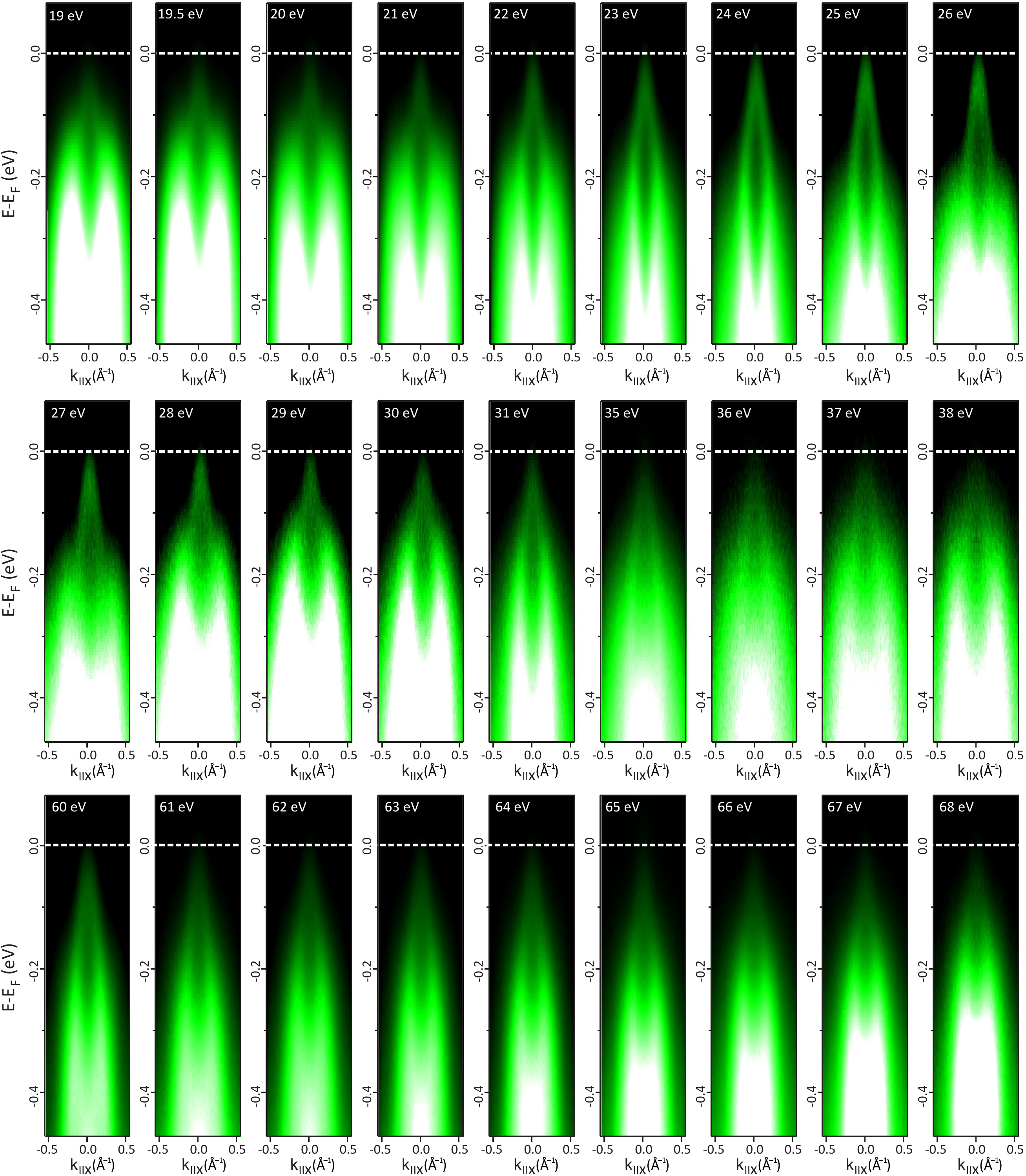}
\vspace{-0.0cm}
\caption{{\bf Photon-energy-dependent ARPES data.} 
Energy-momentum dispersions for photon energies from 19 to 68 eV showing a 2D Dirac cone and the dispersion of 3D bulk states with photon energy. The periodicity in the bulk dispersion demonstrates that the bulk Brillouin zone is probed completely.} 
\label{EDFig4a}
\end{figure}

Figure~\ref{EDFig4a} shows that the Dirac cone surface state has no dispersion with the wave vector perpendicular to the surface plane, in contrast to the bulk valence band. Figure~\ref{EDFig4a} is the input for Fig.~\ref{Fig2}b, main text, with linear interpolation between the 9 photon energies. Table~\ref{tab2} shows the momentum values  in kinetic energy. 

\begin{table}
\vspace{-3cm}
\includegraphics*[scale=0.9]{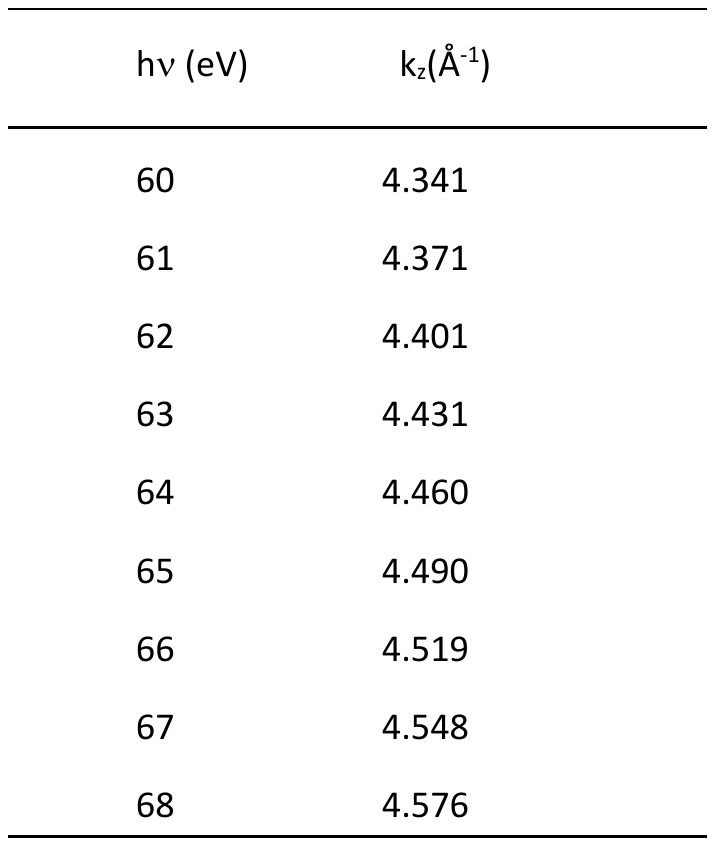}
\vspace{-17cm}
\caption{{\bf Perpendicular momentum.} Determination of the perpendicular 
momentum for the data shown in Fig.~\ref{Fig2}a,b, main text.
Distance from $\Gamma$ to Z corresponds to 0.1327 \AA$^{-1}$. An inner potential of 11.8 eV was used.}
\label{tab2}
\end{table}

\section{Scanning Tunneling Microscopy and Spectroscopy}
\label{sec:S9}
\subsection{Measurement Details}
\label{sec:S9a}
STM measurements were conducted in a home built UHV-STM operating down to 
$T = 4.3$\,K. Cr tips were firstly etched ex-situ and additionally prepared in UHV by field emission on clean W(110). Topography images were recorded in constant-current mode at a tunneling current $I$ and bias voltage $V$ applied to the sample. The \dIdV$(V)$ spectra for scanning tunneling spectroscopy (STS) were recorded \bl{after firstly stabilizing the tip-sample distance at voltage $V_{\rm Stab}=-0.1$\,V and current $I_{\rm Stab}=0.1$\,nA, if not mentioned differently in the captions. Afterwards, the feedback loop was opened and \dIdV\ was recorded} using standard lock-in technique with modulation frequency $f = 1219$\,Hz and amplitude $V_{\rm mod} = 1.4$\,mV while ramping $V$. The spectra were normalized to account for remaining vibrational noise during stabilization by equilibrating the integral between $V_{\rm Stab}$ and $V=0$\,mV.
\begin{figure}
\includegraphics*[scale=0.3]{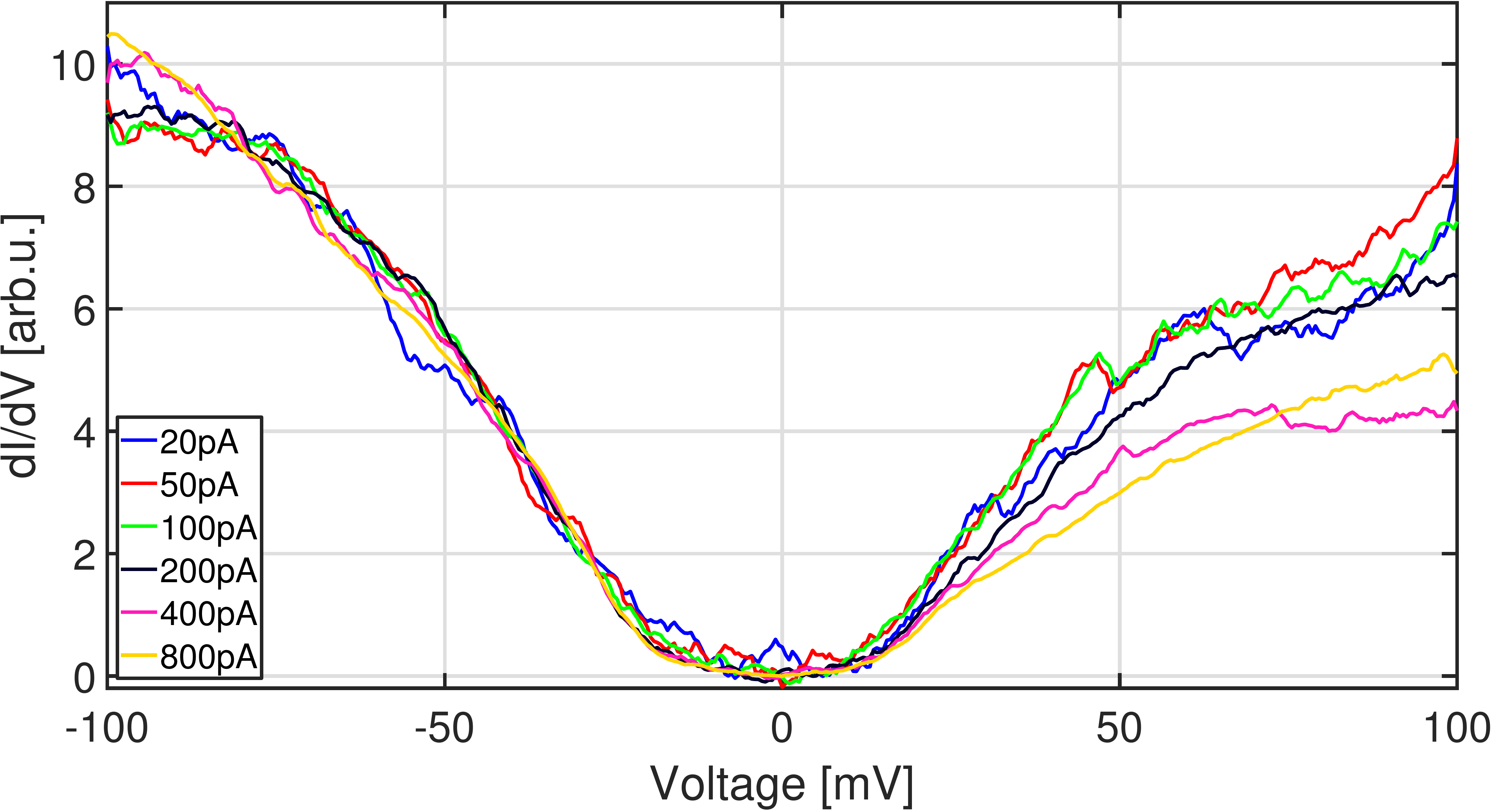}
\vspace{-0.3cm}
\caption{{\bf STS curves recorded at different set points.}
$dI/dV(V)$ curves recorded at the same position after stabilizing the tip 
at $V_{\rm stab}=-100$\,mV and various $I_{\rm stab}$ as marked, $T=4.3$\,K.}
\label{EDFig5a}
\end{figure}
 We crosschecked that $dI/dV$ curves barely depend on the chosen $I_{\rm stab}$  and, hence, on the tip-surface distance (Fig.~\ref{EDFig5a}).

For measurements above the base temperature of 4.3 K, the STM body was exposed to thermal radiation via opening of a radiation shield until a maximum $T \simeq 60$\,K was achieved. Then, the shield was closed, and \dIdV$(V)$ curves were recorded while the sample temperature slowly decreased back to 4.3 K. Measurements at constant $T > 4.3$\,K were performed with partly open shield.
A more frequent stabilization between subsequent $dI/dV$ curves was necessary due to the remaining thermal drift of the tip-sample distance during 
the cooling process. This implies shorter recording times that have been compensated by a more intense averaging of subsequently recorded curves. 

\subsection{Band Gap Determination at 4.3\,K}
\label{sec:S9b}

The band gaps at $T=4.3$\,K were determined as follows. First, the noise level of the $dI/dV$ curves was reduced by averaging $3\times 3$ curves 
covering an area of (1.2\,nm)$^2$.  Subsequently, an averaging of $dI/dV(V)$ across $\pm 2$\,mV in bias direction was employed. Then, we estimated 
the remaining $dI/dV$ noise level as the maximum of $|dI/dV|$  that appears with similar strength positively and negatively. For that purpose, we employed about 50 randomly selected $dI/dV(V)$ curves. Afterwards, the threshold was chosen slightly above the determined noise level.
This threshold is marked in Fig.~\ref{Fig3}a \#1 as dashed line and in Fig.~\ref{Fig3}c as yellow line in the color code bar. 
The voltage width, where the $dI/dV(V)$ spectra stayed below this threshold, defines the measured gap $\Delta$ as used in Fig.~\ref{Fig3}a--c, main text. We crosschecked that $dI/dV$ values below the noise level appeared exclusively close to the determined band gap areas. The resulting gap size  $\Delta$ turned out to barely depend on details of the chosen noise threshold.

\begin{figure}
\includegraphics*[scale=0.9]{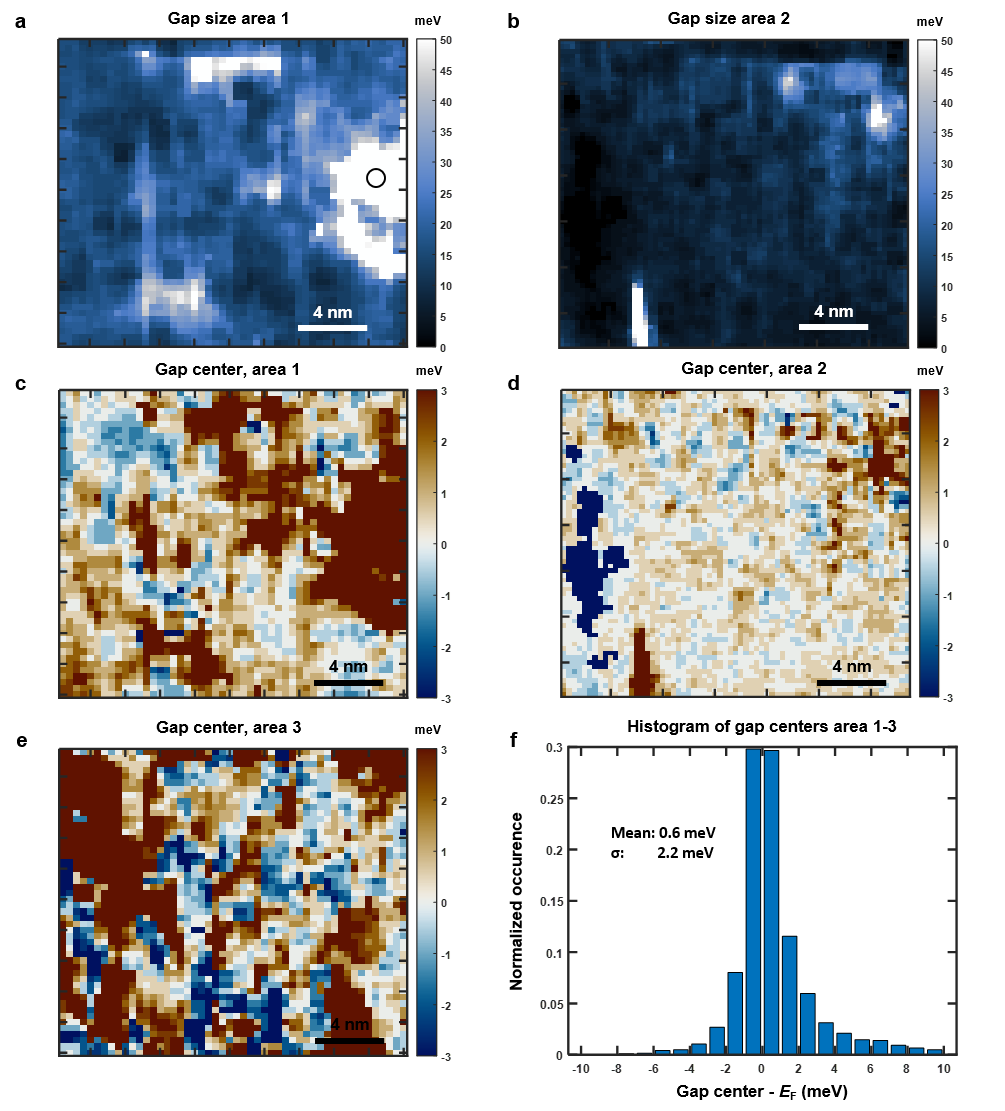}
\vspace{-0.4cm}
\caption{{\bf Maps of gap size $\Delta$ and gap center $E_0$ at 4.3\,K.}
 \bl{(a,b) Spatial maps of the gap size $\Delta(x,y)$ as obtained from $dI/dV(V)$ curves recorded at $T=4.3$\,K. Two additional surface areas named area 1 and area 2 are shown, while area 3 is displayed in Fig.~\ref{Fig3}b, main text. The black circle marks a region, where the determination of a band gap was  not possible (see text).
 \bl{(c)-(e) Spatial maps of the center of the gap $E_0(x,y)$ with respect to the Fermi level \Ef\ for the 3 different areas. (f) Histogram of $E_0$ for all three maps with marked mean and standard deviation $\sigma$.} }}
\label{EDFig6}
\end{figure}

Figure~\ref{EDFig6}a and b display two additional  $\Delta(x,y)$ maps like the ones in  Fig.~\ref{Fig3}b, main text, but recorded on different areas of the sample surface. They exhibit a similar range of spatial fluctuations of $\Delta$ as shown in Fig.~\ref{Fig3}b, main text.
Sometimes, $\Delta$ could not be determined from the $dI/dV(V)$ curves as, e.g., in the bright area marked by a black circle in Fig.~\ref{EDFig6}a.
There, the spectra stayed below the threshold on the positive $V>0$\,mV side up to 100\,mV while a gap edge is observed only on the negative side, 
for an unknown reason. These spectra ($\sim 5$\,\% of all spectra) are discarded from further analysis including the histogram of Fig.~\ref{Fig3}b, main text.
The correlation length $\xi$ of $\Delta(x,y)$ is calculated as FWHM of the correlation function resulting in $\xi\simeq 2$\,nm as given in the main text.

 The central energy within the gap, $E _0$,  is deduced as the arithmetic 
mean of all voltages where the \dIdV\ signal remains below the threshold.
 Figure~\ref{EDFig6}c--f displays maps  $E_0(x,y)$ for the three studied areas as well as a resulting $E_0$ histogram. Favorably, the average of $E_0$ is rather precisely at \Ef\ showing only small spatial fluctuations in the meV range.
 The small discrepancy of the average $E_0\simeq E_{\rm F}$ found by STS to the Dirac point determined by ARPES (20\,meV above \Ef) might be due to the different recording temperatures (4.3\,K vs. 300\,K) or to sample to sample variations.
 
\subsection{$dI/dV(V)$ Curves at Different Temperatures}
\label{sec:S9c}
\begin{figure}
\includegraphics*[scale=0.65]{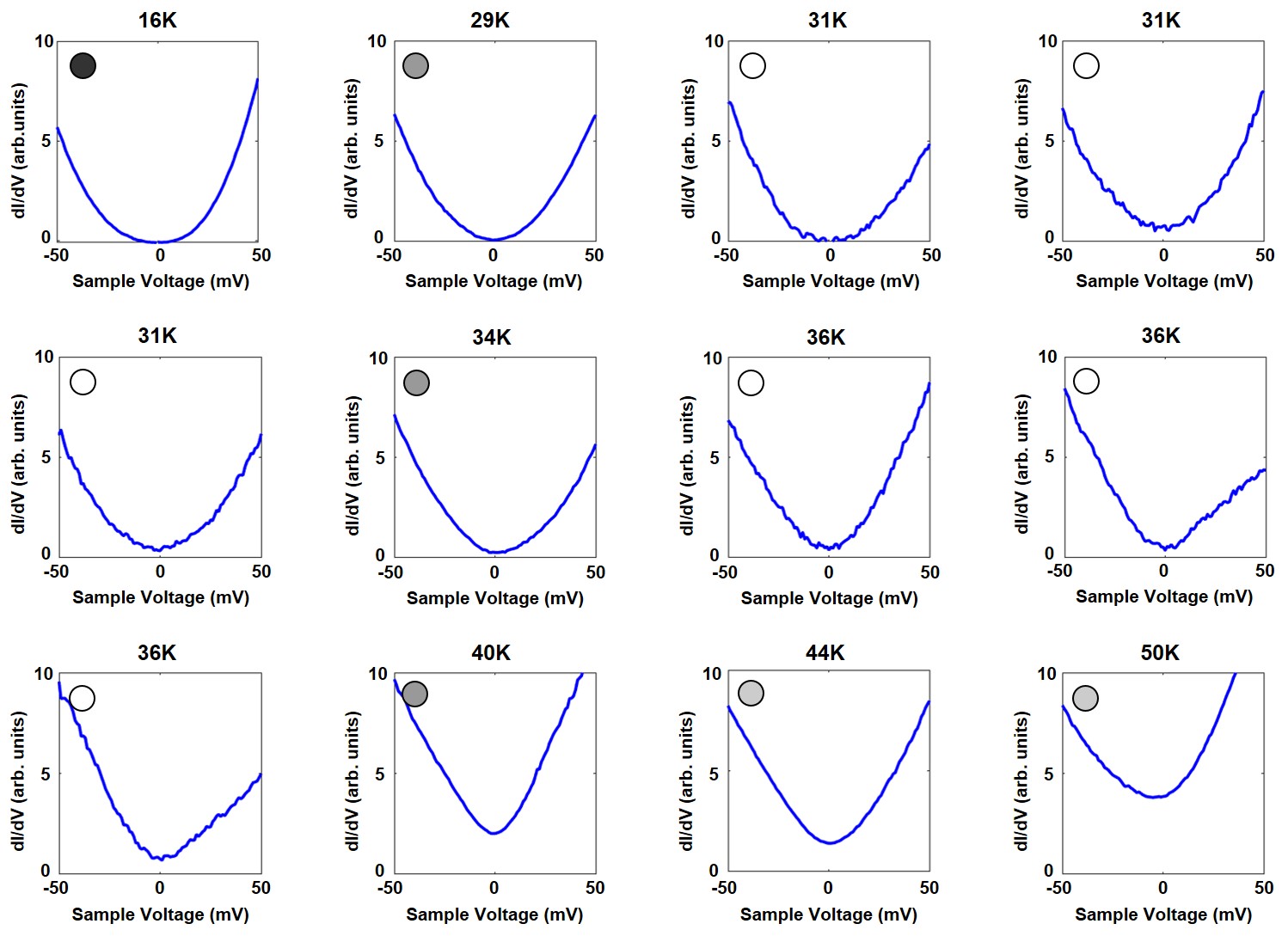}
\vspace{-1cm}
\caption{{\bl{{\bf STS spectra {recorded at different temperatures}.}} 
\bl{Selected \dIdV (V) curves ordered with increasing $T$ as marked on top. The  dot in the upper left is colored identically to the extracted band gap from the same curve as displayed in Fig.~\ref{Fig3}d, main text.}}} 

\label{EDFig7}
\end{figure}

Figure~\ref{EDFig7} displays $dI/dV(V)$ spectra recorded  at various temperatures as used for the band gap evaluation displayed in Fig.~\ref{Fig3}d, main text. An obvious band gap with a voltage region of $dI/dV\simeq 0$\,nS is only found up to about 30\,K. At larger $T$, the curves partly stay close to $dI/dV=0$\,nS around \Ef, but partly strongly deviate from 
$dI/dV=0$\,nS. This varying behaviour requires a more detailed analysis 
to determine $\Delta$, since a small deviation from $dI/dV=0$\,nS might 
also be caused by Fermi level broadening of the tip that  probes a local density of states (LDOS) of the sample with a small band gap. 

\subsection{Determination of Band Gaps at Elevated Temperature} 
\label{sec:S9d}
\begin{figure}
\vspace{-1cm}
\includegraphics*[scale=0.85]{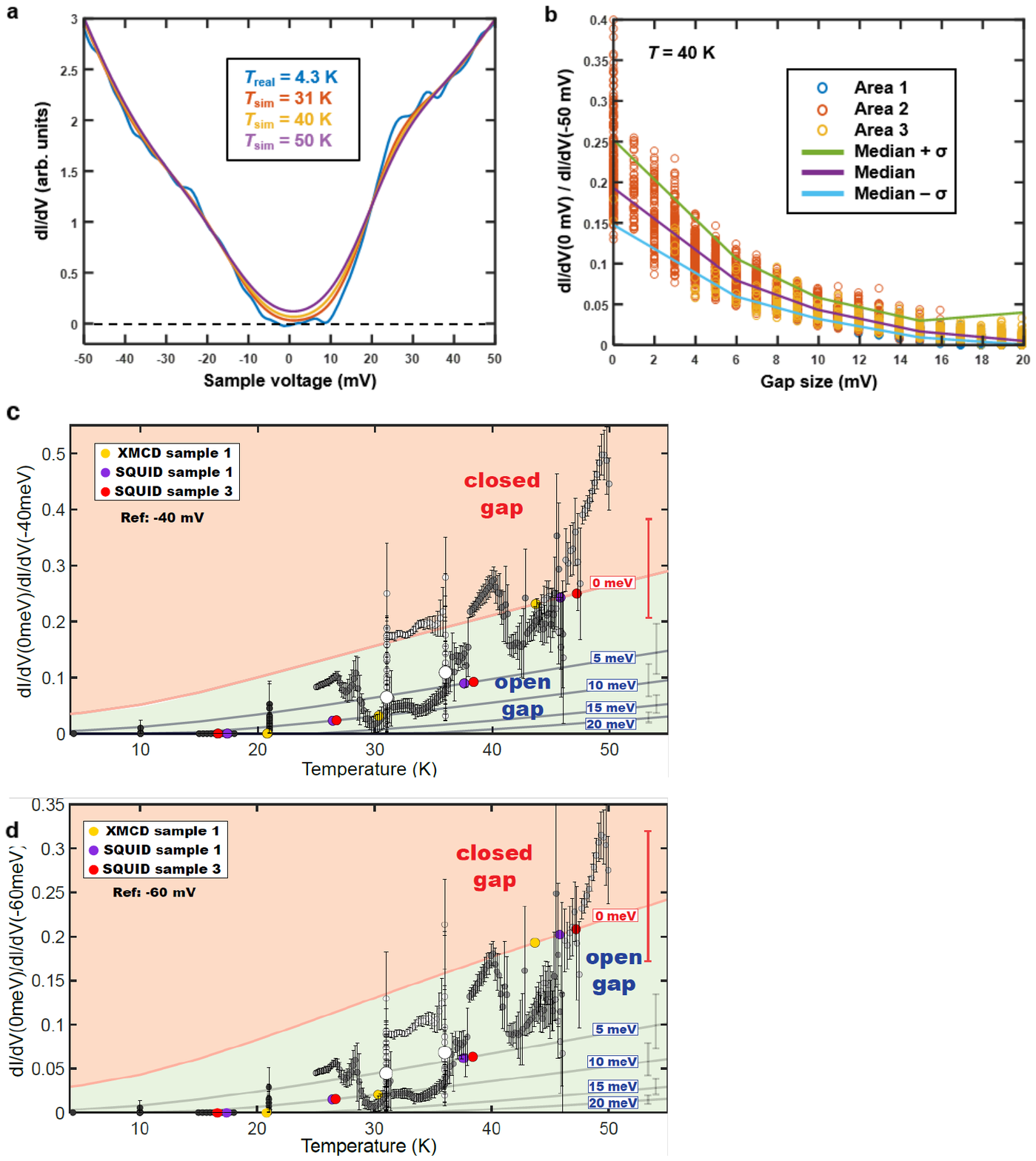}
\vspace{-9.5cm}
\caption{{\bf Gap size determination at elevated $T$.} 
(a) $dI/dV(V)$ spectrum with gap $\Delta = 17$\,meV recorded at 4.3\,K (blue) and the same spectrum after {convolution with the derivative of the Fermi-Dirac} distribution function  for $T=31$\,K ({red}), $T=40$\,K (yellow) and $T=50$\,K (violet). The Fermi level broadening leads to an apparently ungapped 
$dI/dV$ curve, albeit the sample LDOS features a gap of 17\,meV.
 (b) Relation between the measured gap size $\Delta$ of $dI/dV(V)$ curves 
recorded at 4.3\,K and the  ratio $R=[dI/dV(V=0\hspace{1mm} {\rm mV})]/[dI/dV(V=-50\hspace{1mm} {\rm mV})]$ of the same $dI/dV(V)$ curves deduced after convolution with the derivative of the Fermi-Dirac distribution function at 40\,K.  Different circle colors: different areas (Fig.~\ref{EDFig6}). Colored lines: median (violet) surrounded by the standard deviation $\pm\sigma$ (green, blue). (c,d) Same as Fig.~\ref{Fig3}d, main text, but using a different reference voltage $V_{\rm ref}$ to determine $R$. (c) $V_{\rm ref}=40$\,mV, (d) $V_{\rm ref}=60$\,mV.
}
\label{EDFig5}
\end{figure}

 The requirement of a more detailed analysis is
 demonstrated in Fig.~\ref{EDFig5}a. 
 A $dI/dV(V)$ spectrum exhibiting a gap size $\Delta= 17$\,meV  is displayed (blue line) as recorded at 4.3\,K. This spectrum is then convolved with the derivative of the Fermi-Dirac distribution function, proportional to $1/\cosh ^2 (eV/2k_{\rm B}T)$ \cite{Morgenstern03} (other colored lines in Fig.~\ref{EDFig5}a). By this convolution, we mimick the expected appearance of the same spectrum at elevated $T$. Obviously, the previously 
introduced method of gap determination would obtain $\Delta = 0$\,meV for all $T\ge 31$\,K, albeit the LDOS of the sample still exhibits $\Delta 
= 17$\,meV. 
 
In order to deduce the correct $\Delta$ from $dI/dV(V)$ curves recorded at such large $T$, we introduce the ratio $R=[\hbox{\dIdV}(0\hspace{1mm} 
{\rm mV})] / [\hbox{\dIdV}(-50\hspace{1mm} {\rm mV})]$ that turns out to be monotonously anticorrelated with the gap size $\Delta$. This is demonstrated in Fig.~\ref{EDFig5}b for $T=40$\,K. All $dI/dV(V)$ curves recorded at 4.3\,K (area 1--3, 8000 curves) are convoluted with the derivative 
of the Fermi-Dirac distribution of 40\,K (as in Fig.~\ref{EDFig5}a, yellow curve) before $R$ is determined. Subsequently, each $R$ is related to its corresponding $\Delta$ (same $dI/dV(V)$ curve) as deduced via the method described in subsection~\ref{sec:S9b}. The anticorrelation of $R$ and $\Delta$ appears in
Fig.~\ref{EDFig5}b and is similarly found for all $T=25-50$\,K (not shown). 
We used the resulting median of the simulated $R$ values  (violet line in 
Fig.~\ref{EDFig5}b) to deduce $\Delta$ from a measured $R$ for each $dI/dV(V)$ curve recorded at elevated $T$. The required simulated $R(T)$ curves for constant $\Delta$ are displayed as full lines in Fig.~\ref{Fig3}d, main text, for the sake of comparison with the $R$ values of the measured 
\dIdV\ curves.
The error of $R(\Delta)$ at given $T$ is taken as the  $2\sigma$ width of 
the simulated $R(\Delta)$, marked by colored lines in Fig.~\ref{EDFig5}b. 
For gap sizes below 10\,meV, one observes an error of 2--4\,meV only. 
Overall, the $R(\Delta)$ relation reaches a relative accuracy for $\Delta$ determination of $\pm 30$\,\%, as long as $\Delta$ remains below 20 meV. This error is displayed exemplarily on the very right of Fig.~\ref{Fig3}d, main text.

In addition, we determined errors for the measured $R$ values. They are deduced as the standard deviation of subsequently recorded 25 \dIdV$(V)$ data sets (averaged from 10 subsequent \dIdV\ curves each and smoothed by box averaging of width 3\,mV). The measurement errors are displayed as error bars at the data points in Fig.~\ref{Fig3}d, main text (circles). Interestingly, these measurement errors increase significantly  around $T\simeq 45$\,K, i.e., close to \Tc, where they get as large as 10\,meV. 
Since the spatial drift during recording of the 25 data sets is below 1\,nm, as deduced by comparing the long term development of \dIdV\ data during cooling {to the} spatial fluctuations recorded at 4.3\,K, these relatively large errors cannot be caused by a lateral drift of the tip with respect to the sample only. Likely, they are caused by enhanced temporal fluctuations of the gap close to \Tc\ during recording of the 25 data sets. Thus, the increased error bars around \Tc\ corroborate the relation of the band gap evolution to the magnetic properties additionally. 

Error bars of the spatially averaged data points in Fig.~\ref{Fig3}d, main text, indicate the width of the $R$ histograms obtained at the corresponding $T$.

The reference voltage $V_{\rm ref}=-50$\,mV used to determine $R$ is chosen such that it is not influenced by temperature ($|{eV_{\rm ref}}|\gg 5 k_{\rm B}T$) or gap size ($|{eV_{\rm ref}}|\gg \max({\Delta})/2$) and not influenced by spectroscopic features that might spatially vary due to disorder. We crosschecked that 
the exact value of $V_{\rm ref}$ barely influences the deduced $\Delta$. 
This is demonstrated in Fig.~\ref{EDFig5}c--d displaying the same data set as Fig.~\ref{Fig3}d, main text, but using two different $V_{\rm ref}$.
 
\section{{Electric Transport Measurements}}
\label{sec:S10}
\bl{Transport measurements were performed in the van der Pauw geometry with applied magnetic fields ranging from -3\,T to +3\,T and oriented parallel to the rhombohedral axis of the epilayers. A mini cryogen-free system 
was employed for the magnetotransport investigations at temperatures between 2\,K and 300\, K. For temperatures below about 50\, K an anomalous Hall effect appears in the transport data that becomes hysteretic for magnetic fields of less than about  1\, T  such as the sheet resistance of the 
samples (compare Fig.~\ref{Fig1a}e, main text).  The carrier concentrations were determined from the Hall resistance at sufficiently high magnetic 
fields, i.e., well above 1\,T where the sample magnetization was largely saturated. The resulting hole concentrations of the MnSb$_2$Te$_4$ samples range from $(1-3) \cdot 10^{20}\mathrm{cm}^{-3}$, similar to values from the literature \cite{McQueeneyPRB19,ChenPRM20}.}

\section{Density Functional Theory Calculations}
\label{sec:S11}
\subsection{Details of the Calculations}
\label{sec:S11a}
The electronic and magnetic structures were calculated by two different methods based on density functional theory (DFT).
Most of the bulk-type calculations (\MM{Figs.~\ref{Fig4}f--k, main text, Fig.~\ref{EDFig8}, Fig.~\ref{EDFig9}a,b}) used a  Green function method within the multiple scattering theory  \cite{Gyorffy1973,Geilhufe2015}. To 
describe both localization and interaction of the Mn $3d$ orbitals appropriately, Coulomb $U$ values  of 3--5\,eV were employed within a \bl{GGA+$U$} approach
  \cite{Anisimov1991}. During these calculations,
we confirmed that different topological phases of \MnSbSbTe4\ arise, if one uses either GGA+$U$ or LDA+$U$, indicating that LDA+$U$ is not sufficient.
To account for antiferromagnetic configurations of \MnSbSbTe4, a double unit cell consisting of two septuple layers was used such as for Fig.~\ref{Fig4}k, main text, and Fig.~\ref{EDFig8}a and c. In these structures, the Mn atoms couple ferromagnetically within the Mn planes and antiferromagnetically between neighboring Mn layers in adajacent septuple layers. For 
the determination of $T_{\rm N}$ and $T_{\rm  C}$, exchange   constants $J_{ij}$ were obtained by mapping the DFT calculations onto a classical Heisenberg model  \cite{Liechtenstein1987}. 

Different types of disorder were treated within a coherent  potential approximation (CPA) \cite{Soven1967,Gyorffy1972}.   
Chemical disorder is modelled by mixing various atomic species on the same atomic site (substitutional alloys). The elemental unit cell was used in this case such that only ferromagnetic order could be described. We simulated three types of chemical disorder, namely site exchange, i.e., placing as much Mn on Sb sites as Sb on Mn sites, Sb excess, i.e., additional, substitutional Sb in the Mn layers, and Mn excess, i.e., additional, substitutional Mn in the Sb layers.  
Electron (hole) doping was also modeled by CPA via mixing Te (Sb) vacancies, that were simulated as empty spheres, with Te (Sb) atoms on the same site.
Magnetic moment disorder was modeled by CPA via mixing two Mn atoms with opposite magnetic moment on the same atomic site. This approach has been proven to be very successful in mimicking spin moment fluctuations, e.g., 
to account for elevated temperatures. The $50 : 50$ mixing represents the 
paramagnetic state, while smaller spin   fluctuations are described by, e.g., $95 :  5$ or   $98:  2$ mixings.  

The calculations in Fig.~\ref{EDFig9}c--f were  performed with the full-potential linearized augmented plane-wave method as implemented in the FLEUR code. Also here, GGA \cite{Perdew.prl1996} with a Hubbard $U$ correction using $U=6$ eV and $J=0.54$ eV was used and spin-orbit coupling was included self-consistently in the non-collinear calculations \cite{Kurz04}.
Thin film calculations were also performed with the FLEUR code (\MM{Fig.~\ref{Fig4}c, main text, and Fig.~\ref{EDFig10}b--e}). These calculations are restricted to relatively small and simple unit cells in order to retain high enough accuracy to derive the Dirac cone dispersion and the size of its magnetic gap by DFT.
Instead, the surface band structure for the antiferromagnetic ground state in Fig.~\ref{Fig4}d--e, main text, for the ferromagnetic ground state in Fig.~\ref{Fig4}b, main text, and for the mixed state in Fig.~\ref{EDFig10}a was  calculated  by the projector augmented-wave method \cite{ref21} 
using the VASP code \cite{ref22,ref23}.  
The exchange-correlation energy was treated using the GGA \cite{Perdew.prl1996}. The Hamiltonian contained scalar relativistic corrections and the 
spin-orbit coupling was taken into account by the second variation  method \cite{Koelling.jpc1977}. In order to describe the van der Waals interactions, we made use of the DFT-D3 \cite{Grimme.jcp2010, Grimme.jcc2011} approach.
The Mn $3d$  states were treated employing the GGA$+U$ approximation \cite{Anisimov1991} within the Dudarev scheme \cite{Dudarev.prb1998}. The $U_\text{eff}=U-J$ value for the Mn 3$d$ states
was chosen to 5.34~eV as in previous works \cite{EremeevJAC17,Otrokov2DMat,OtrokovAFMTI18,OtrokovPRL19}.

For all three methods, the crystal structure of  ideal \MnSbSbTe4\ was fully optimized to obtain the equilibrium lattice parameters, namely
cell volume, $c/a$ ratio as well as atomic positions. Using this structure yields the antiferromagnetic topological insulator state for \MnSbSbTe4\ within both the projector augmented-wave and full-potential linearized augmented plane-wave methods (VASP and FLEUR, respectively). For the Green function method both the experimental crystal structure as determined by XRD (Fig.~\ref{EDFig1}) and the theoretically optimized one  result in the antiferromagnetic topological insulator  phase.

In order to visualize the topological character within bulk-type band structure calculations,
we analyze the spectral function difference between anion and cation contributions of each state: $A^i_k(E) = A_k^{\rm anion}(E)-A_k^{\rm cation}(E)$.  The resulting $A^i_k(E)$ are displayed as color code
in Fig.~\ref{Fig4}f--k, main text, Fig.~\ref{EDFig8}, and Fig.~\ref{EDFig9}a,b with red color for $A_k^i(E) > 0 $ and blue color for $A_k^i(E) < 0 
$. Band inversion can, hence, be deduced from a mutually changing color within adjacent bands. 

\subsection{Magnetic Ground State for Different Disorder Configurations}
\label{sec:S11b}

\begin{table}
\vspace{-2cm}
\includegraphics*[scale=0.9]{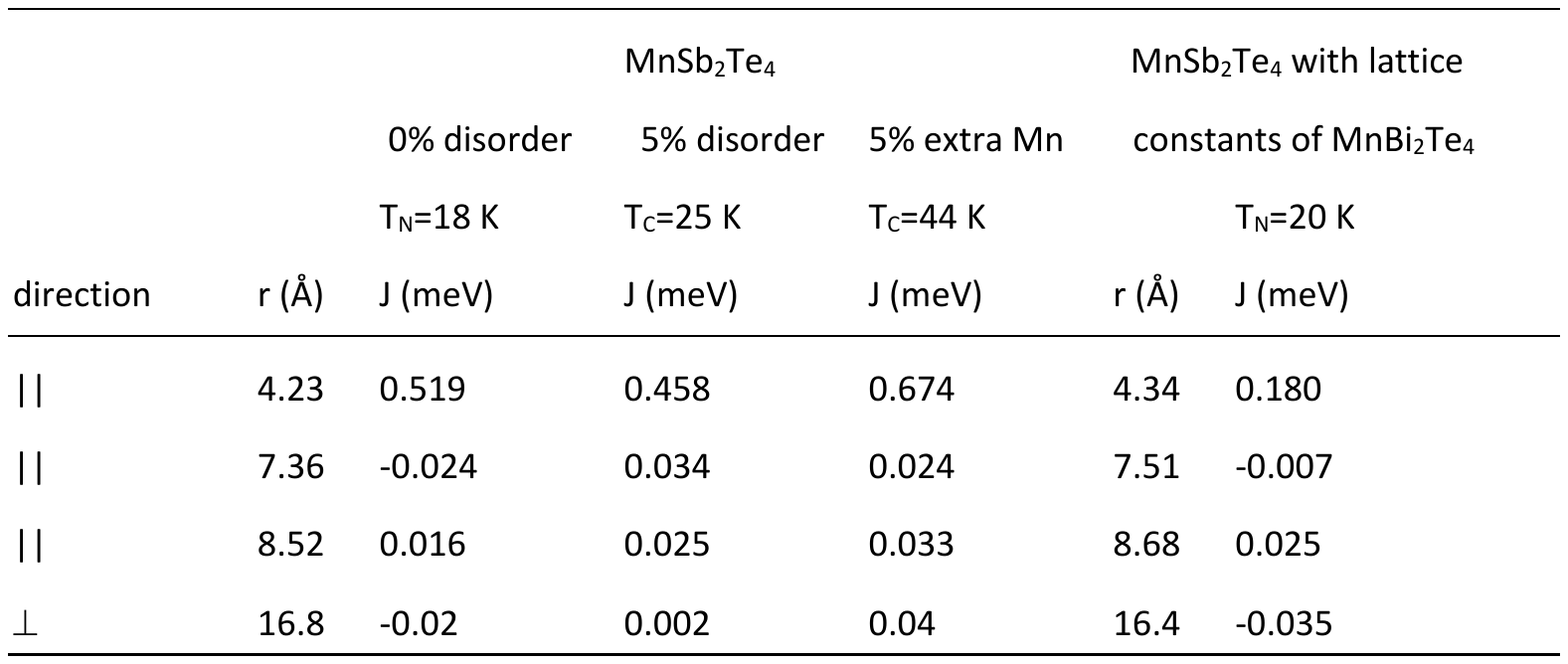}
\vspace{-20cm}
\caption{{\bf Magnetic structure, critical temperature, and Mn exchange integrals of \MnSbSbTe4\ for four different crystal configurations (DFT).} 
The exchange integrals $J$ for the four smallest distances $r$ between the contributing Mn ions are additionally marked by $\parallel$ ($\perp$) for intralayer (interlayer) Mn pairs. The model with 0\%\ disorder refers to ideal \MnSbSbTe4\ with all Mn atoms in the central plane of the septuple layers at the experimental lattice constants $a=4.23$ \bl{\AA, $c=40.98$ \AA\ (XRD). In the second model (5\%\ disorder), a Mn-Sb site exchange leads to an occupancy of {95\%\ Mn and 5\%\ Sb} in the central lattice plane and {2.5\%\ Mn and 97.5\%\ Sb} in each of the two outer cationic 
lattice planes of the septuple layer. In the third case, 2.5\%\ extra Mn was substitutionally introduced in the Sb layers, such that the outer cationic planes contain 2.5\%\ Mn and 97.5\%\ Sb, while a 100\%\ Mn occupancy appears in the central layer.
The  fourth model features ideal \MnSbSbTe4\ with 0\%\ disorder, but laterally 
expanded to the lattice constants of \MnBiBiTe4, i.e., $a=4.34$ \AA, $c 
= 40.89$ \AA.}}
\label{tab3}
\end{table}

Table~\ref{tab3} summarizes the magnetic properties for four different crystallographic configurations as obtained by bulk-type DFT calculations. The ideal \MnSbSbTe4\ septuple layer configuration is antiferromagnetic with low $T_{\rm N}=18$\,K (third row). 
To probe the role of the lattice constant for the magnetic properties, we 
slightly expanded the lattice within the planes
towards the lattice constant of \MnBiBiTe4\ (last row). This led to a reduced nearest neighbor exchange constant $J$, but barely to a change in $T_{\rm N}$. Introducing 5\,\% Mn-Sb site exchange, in line with the XRD and STM data, turned the interlayer coupling ferromagnetic, but still with a \Tc$=25$\,K only, i.e., significantly lower than the experimental value (fourth row).
However, exchanging 2.5\,\% of the Sb within the Sb layers by Mn without changing the Mn layers revealed \Tc$=44$\,K, very close to the experimental value (fifth row).
Hence, we conclude, in line with RBS, STM and XRD data, that an excess Mn 
in combination with an Sb deficiency is responsible for the ferromagnetic 
behaviour of \MnSbSbTe4\ with high \Tc\ via Mn substitution in the Sb layers.

\subsection{Topological Properties for Different Disorder Configurations}
\label{sec:S11c}
\begin{figure}
\includegraphics*[scale=0.66]{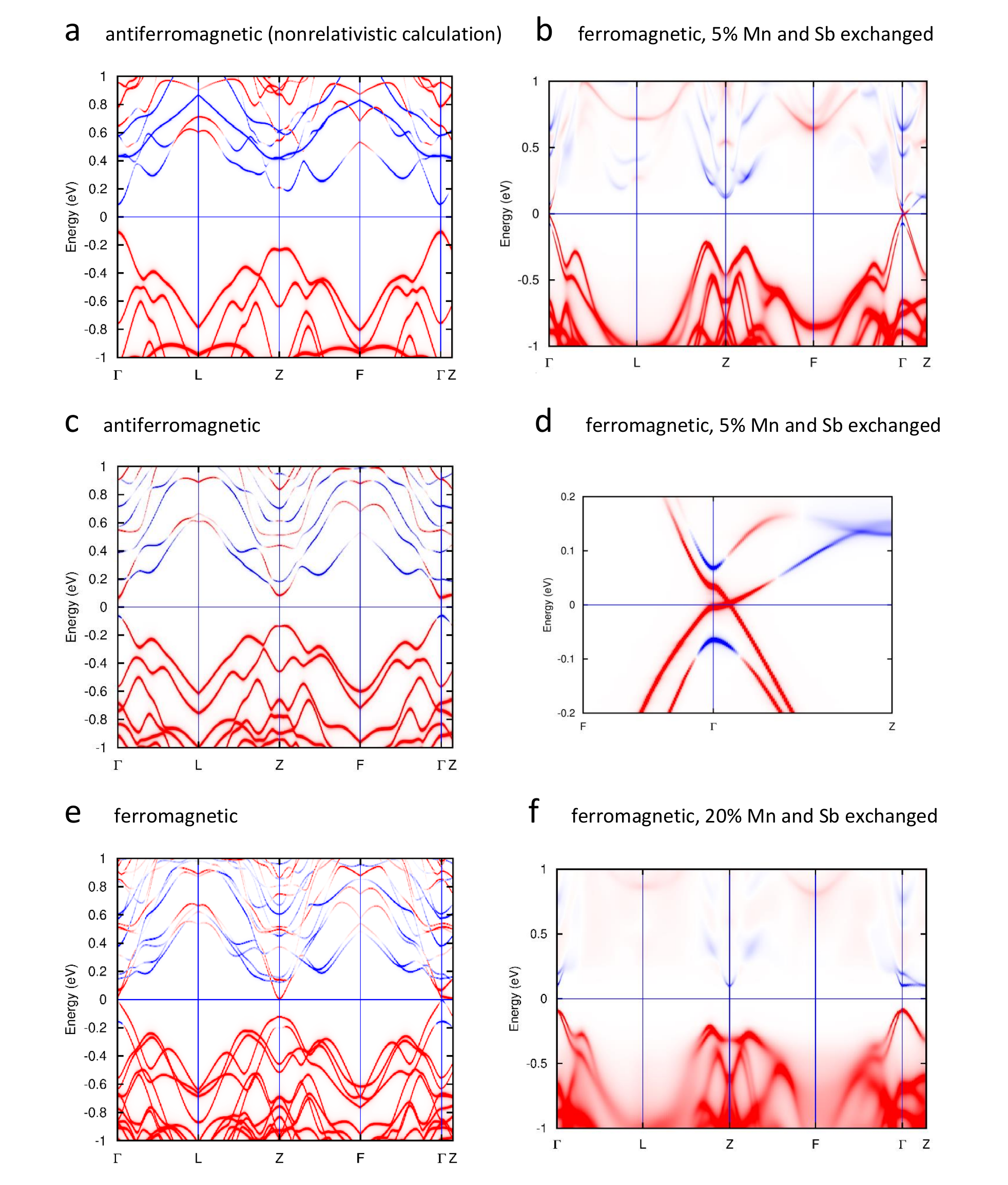}
\vspace{-0.5cm}
\caption{{\bl{{\bf Band Structure at Different Strength of Mn-Sb Site Exchange (DFT)}.}}
\bl{Color represents the difference between anionic and cationic spectral 
function for each state, red: more anionic, blue: more cationic.  (a) Antiferromagnetic, ideal \MnSbSbTe4\ in a nonrelativistic (more precisely scalar relativistic) calculation. No band inversion occurs. (c) Antiferromagnetic, ideal \MnSbSbTe4\ in a fully relativistic calculation yielding a band inversion. Only one inversion appears in the Brillouin zone (at $\Gamma$) evidencing a $Z_2$ topological insulator (also shown as Fig.~\ref{Fig4}k, main text, for the ferromagnetic unit cell.). (e) Same as (c) but for ferromagnetically ordered, ideal  \MnSbSbTe4\ }  {with perfect out-of-plane alignment of Mn moments}. A 3D Weyl semimetal occurs. The calculation used the antiferromagnetic unit cell for better comparison with (c). (b, d, f) Ferromagnetic  \MnSbSbTe4\ {with different Mn-Sb site exchange as labeled (ferromagnetic unit cell). In (b) and (d), the Weyl point is preserved, while (f)  shows a topologically trivial band gap.}} 
\label{EDFig8}
\end{figure}

Figure~\ref{EDFig8} shows bulk band structure calculations for antiferromagnetic \MnSbSbTe4\ with ideal stoichiometric order  (a,c), ideal, ferromagnetic \MnSbSbTe4\ (e)  and ferromagnetic \MnSbSbTe4\ with increasing Mn-Sb site exchange (b,d,f). Antiferromagnetic \MnSbSbTe4\ is a topological 
insulator with inverted band gap at $\Gamma$, if spin-orbit coupling is considered (c). 
This is contrary to previous calculations that have suggested antiferromagnetic  \MnSbSbTe4\ to be trivial \cite{ZhangDPRL19,ChenBo19,Lei20,LiuY20}. We assume that the topological insulator state of antiferromagnetic \MnSbSbTe4\ was missed  because structural optimization 
was either not performed in favor of experimental lattice constants  \cite{ChenBo19,LiuY20} or performed without
 van der Waals forces \cite{ZhangDPRL19,Lei20} which both gave by $\sim3$\%\ larger $c$ parameters than in the structurally optimized equilibrium lattice. 
Ferromagnetic \MnSbSbTe4\, instead, is a Weyl semimetal without band gap that remains a Weyl semimetal for moderate Mn-Sb site exchange (Fig.~\ref{EDFig8}b,d), but becomes topologically trivial at large site exchange (Fig.~\ref{EDFig8}f).
Hence, site exchange alone does not reveal the experimentally observed ferromagnetic topological insulator in contrast to the magnetic disorder as 
presented in  Fig.~\ref{Fig4}g--j, main text.

\begin{figure}
\includegraphics*[scale=0.72]{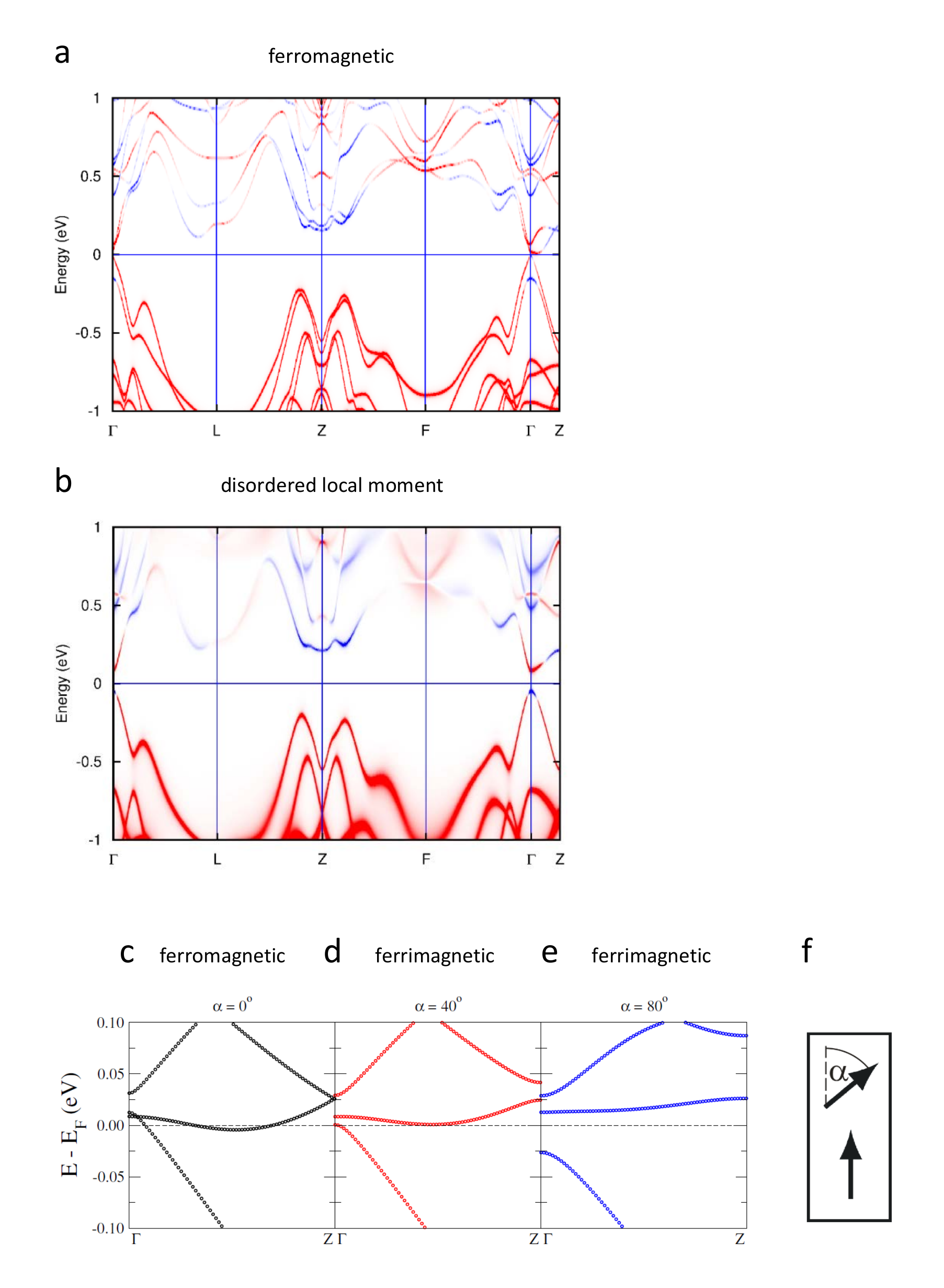}
\vspace{-0.9cm}
\caption{\bl{{\bf Inverted band gap by magnetic disorder (DFT).}} 
(a) Bulk band structure of ideal, ferromagnetic  \MnSbSbTe4\  with Mn moments perpendicular to the septuple layers (single layer unit cell). (Part 
of this figure is shown as Fig.~\ref{Fig4}f, main text.)  (b) Same as (a), but with disordered local magnetic moments (50\%\ spin up, 50\%\ spin down). An inverted band gap appears at $\Gamma$. (Part of this figure is shown as Fig.~\ref{Fig4}j, main text.)
(c--e) Alternative demonstration of gap opening at the Weyl point (two septuple layer unit cell): one layer exhibits a collinear out-of-plane ferromagnetic order, while the other is also collinear, but canted relative to the first one by an angle $\alpha$ as marked. (f) Vector model of the collinearly canted two septuple layers.} 
\label{EDFig9}
\end{figure}

Figure~\ref{EDFig9} corroborates the gap opening by magnetic disorder. Figure~\ref{EDFig9}b displays the complete band structure at maximum spin mixture (50\,\% spin-up, 50\,\% spin down) for each Mn lattice site as partially presented in Fig.~\ref{Fig4}j, main text. An inverted band gap of about 100\,meV is found at $\Gamma$. Note that the purely ferromagnetic phase in Fig.~\ref{EDFig9}a is calculated for a unit cell of a single septuple layer only and, hence, differs from Fig.~\ref{EDFig8}e employing a unit cell with two septuple layers, due to backfolding.

 The  Weyl point observed for the ferromagnetic \MnSbSbTe4\ can also be opened by rotating the spins of adjacent Mn layers, while keeping a collinear spin order within each layer (Fig.~\ref{EDFig9}c--f).
Thus, magnetic disorder renders a dominantly ferromagnetic \MnSbSbTe4\ a topological insulator as already discussed for Fig.~\ref{Fig4}g--j, main text.

\begin{figure}
\includegraphics*[scale=0.66]{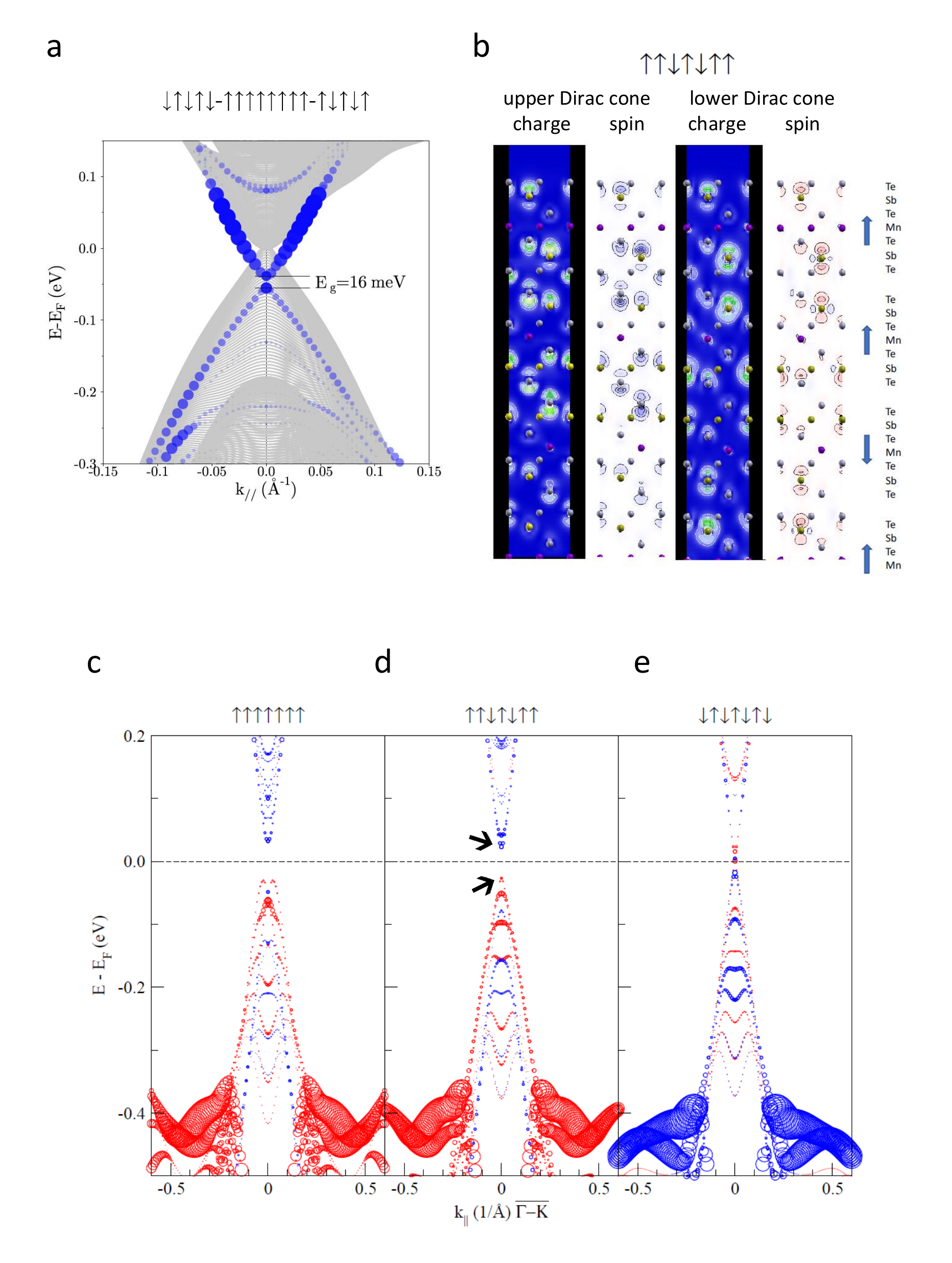}
\vspace{-1cm}
\caption{{\bf Influence of magnetic structure on  topological surface states (DFT).} 
\bl{(a) Band structure for a ferromagnetic bulk
surrounded by antiferromagnetic surface layers as marked on top. Blue dots are surface states with dot diameter marking their strength at the surface septuple layer. A gap appears in the topological surface state. (b) Charge density and spin density of the states at the edges of the magnetically induced band gap in (d) displayed as thin film cross section. The magnetic configuration is marked on top. 
(c--e) Band structure for various magnetic configurations each marked  on 
top. Colors denote the spin density in the out-of-plane direction. (c) Ferromagnetic order with out-of-plane anisotropy. (d) Two ferromagnetic septuple layers on both sides of three antiferromagnetic layers. Black arrows mark the states shown in (b). (e) Perfect antiferromagnetic  order as in Fig.~\ref{Fig4}d,e, main text.} }
\label{EDFig10}
\end{figure}
 
Figure~\ref{EDFig10} shows the band structures of slab calculations for different magnetic disorder configurations, such that surface states are captured. 
They are performed, e.g., for a combination of ferromagnetic interior layers surrounded by a few antiferromagnetic layers on top and bottom (Fig.~\ref{EDFig10}a). This configuration reveals a Dirac-type surface  state with a gap around the Dirac point of 16\,meV,  very close to the average gap size observed by STS. The opposite configuration with antiferromagnetic interior surrounded by ferromagnetic surfaces also exhibits a gapped Dirac cone, here with 40\,meV gap size, that might be enhanced by the thickness of this slab of only 7 septuple layers (Fig.~\ref{EDFig10}d). Indeed, the spin-polarized states at the gap edge penetrate about 3 septuple layers into the bulk of the thin film (Fig.~\ref{EDFig10}b).
The out-of-plane spin polarization near the Dirac point amounts to $\sim60$\%, nicely matching the experimentally found out-of-plane spin polarization in spin-resolved ARPES (Fig.~\ref{Fig2}j, main text). The latter amounts to $\sim 25$\,\% at 30\,K in line with the reduced magnetization at this elevated $T$ (Fig.~\ref{Fig1a}d,f, main text). 
Note that the band structure in Fig.~\ref{EDFig10}d  also features an exchange splitting of bulk bands as visible by the different colors around $-0.2$\,eV.

The pure antiferromagnetic configuration (Fig.~\ref{EDFig10}e) shows a small band gap of the topological surface state as well, while the pure ferromagnetic order leads to a gapped Weyl cone (Fig.~\ref{EDFig10}c), likely being an artifact of the finite slab size of 7 septuple layers only. 

\subsection{Influence of Charge Doping on the Magnetic Interactions}
\label{sec:S11d}
Finally, the influence of charge doping on the magnetic properties of \MnSbSbTe4\ was studied.
For undoped, ideally stacked \MnSbSbTe4, the leading magnetic interaction 
between the septuple layers is of a superexchange type, which is responsible for the antiferromagnetic interlayer coupling. As shown above, chemical disorder by exchanging (replacing) Mn with (by) Sb changes the magnetic order to ferromagnetic. The presence of Mn defects could induce an additional Ruderman-Kittel-Kasuya-Yoshida (RKKY) interaction that might influence the magnetic order, if mobile charges are present. 
However, our calculations show only minor changes of exchange constants and the magnetic transition temperatures upon charge doping. 
For n-type doping by 0.2\%\ Te vacancies, \Ef\ shifts by 0.29 eV increasing \Tc\ by 0.9 K only.
For p-type doping by 0.2\%\ Sb vacancies, \Ef\ shifts by $-0.08$ eV increasing \Tc\ by 1.6 K.
This indicates that the RKKY interaction is negligible in \MnSbSbTe4.


\medskip

%


\begin{thebibliography}{10}
\providecommand{\url}[1]{\texttt{#1}}
\providecommand{\urlprefix}{URL }

\bibitem{Onoda03}
M.~Onoda, N.~Nagaosa,
\newblock \emph{Phys. Rev. Lett.} \textbf{2003}, \emph{90}, 20 206601.

\bibitem{CXLiu08}
C.-X. Liu, X.-L. Qi, X.~Dai, Z.~Fang, S.-C. Zhang,
\newblock \emph{Phys. Rev. Lett.} \textbf{2008}, \emph{101}, 14 146802.

\bibitem{YuScience10}
R.~Yu, W.~Zhang, H.-J. Zhang, S.-C. Zhang, X.~Dai, Z.~Fang,
\newblock \emph{Science} \textbf{2010}, \emph{329}, 5987 61.

\bibitem{Qiao10}
Z.~Qiao, S.~A. Yang, W.~Feng, W.-K. Tse, J.~Ding, Y.~Yao, J.~Wang, Q.~Niu,
\newblock \emph{Phys. Rev. B} \textbf{2010}, \emph{82}, 16 161414.

\bibitem{Tokura19}
Y.~Tokura, K.~Yasuda, A.~Tsukazaki,
\newblock \emph{Nat. Rev. Phys.} \textbf{2019}, \emph{1}, 2 126.

\bibitem{Chang13}
C.-Z. Chang, J.~Zhang, X.~Feng, J.~Shen, Z.~Zhang, M.~Guo, K.~Li, Y.~Ou,
  P.~Wei, L.-L. Wang, Z.-Q. Ji, Y.~Feng, S.~Ji, X.~Chen, J.~Jia, X.~Dai,
  Z.~Fang, S.-C. Zhang, K.~He, Y.~Wang, L.~Lu, X.-C. Ma, Q.-K. Xue,
\newblock \emph{Science} \textbf{2013}, \emph{340}, 6129 167.

\bibitem{CheckelskyNP14}
J.~G. Checkelsky, R.~Yoshimi, A.~Tsukazaki, K.~S. Takahashi, Y.~Kozuka,
  J.~Falson, M.~Kawasaki, Y.~Tokura,
\newblock \emph{Nat. Phys.} \textbf{2014}, \emph{10}, 10 731.

\bibitem{KouPRL14}
X.~Kou, S.-T. Guo, Y.~Fan, L.~Pan, M.~Lang, Y.~Jiang, Q.~Shao, T.~Nie,
  K.~Murata, J.~Tang, Y.~Wang, L.~He, T.-K. Lee, W.-L. Lee, K.~L. Wang,
\newblock \emph{Phys. Rev. Lett.} \textbf{2014}, \emph{113}, 13 137201.

\bibitem{BestwickPRL15}
A.~Bestwick, E.~Fox, X.~Kou, L.~Pan, K.~L. Wang, D.~Goldhaber-Gordon,
\newblock \emph{Phys. Rev. Lett.} \textbf{2015}, \emph{114}, 18 187201.

\bibitem{Kandala2015}
A.~Kandala, A.~Richardella, S.~Kempinger, C.-X. Liu, N.~Samarth,
\newblock \emph{Nat. Commun.} \textbf{2015}, \emph{6}, 1 7434.

\bibitem{ChangCZNM15}
C.-Z. Chang, W.~Zhao, D.~Y. Kim, H.~Zhang, B.~A. Assaf, D.~Heiman, S.-C. Zhang,
  C.~Liu, M.~H.~W. Chan, J.~S. Moodera,
\newblock \emph{Nat. Mater.} \textbf{2015}, \emph{14}, 5 473.

\bibitem{GrauerPRB15}
S.~Grauer, S.~Schreyeck, M.~Winnerlein, K.~Brunner, C.~Gould, L.~W. Molenkamp,
\newblock \emph{Phys. Rev. B} \textbf{2015}, \emph{92}, 20 201304.

\bibitem{FoxPRB18}
E.~J. Fox, I.~T. Rosen, Y.~Yang, G.~R. Jones, R.~E. Elmquist, X.~Kou, L.~Pan,
  K.~L. Wang, D.~Goldhaber-Gordon,
\newblock \emph{Phys. Rev. B} \textbf{2018}, \emph{98}, 7 075145.

\bibitem{GoetzAPL18}
M.~G\"{o}tz, K.~M. Fijalkowski, E.~Pesel, M.~Hartl, S.~Schreyeck,
  M.~Winnerlein, S.~Grauer, H.~Scherer, K.~Brunner, C.~Gould, F.~J. Ahlers,
  L.~W. Molenkamp,
\newblock \emph{Appl. Phys. Lett.} \textbf{2018}, \emph{112}, 7 072102.

\bibitem{YeNatComm2015}
M.~Ye, W.~Li, S.~Zhu, Y.~Takeda, Y.~Saitoh, J.~Wang, H.~Pan, M.~Nurmamat,
  K.~Sumida, F.~Ji, Z.~Liu, H.~Yang, Z.~Liu, D.~Shen, A.~Kimura, S.~Qiao,
  X.~Xie,
\newblock \emph{Nat. Commun.} \textbf{2015}, \emph{6}, 1 8913.

\bibitem{LiMPRL15}
M.~Li, C.-Z. Chang, L.~Wu, J.~Tao, W.~Zhao, M.~H. Chan, J.~S. Moodera, J.~Li,
  Y.~Zhu,
\newblock \emph{Phys. Rev. Lett.} \textbf{2015}, \emph{114}, 14 146802.

\bibitem{MogiAPL15}
M.~Mogi, R.~Yoshimi, A.~Tsukazaki, K.~Yasuda, Y.~Kozuka, K.~S. Takahashi,
  M.~Kawasaki, Y.~Tokura,
\newblock \emph{Appl. Phys. Lett.} \textbf{2015}, \emph{107}, 18 182401.

\bibitem{DengNatPhys2020}
H.~Deng, Z.~Chen, A.~Wo{\l}o{\'{s}}, M.~Konczykowski, K.~Sobczak, J.~Sitnicka,
  I.~V. Fedorchenko, J.~Borysiuk, T.~Heider, {\L}.~Pluci{\'{n}}ski, K.~Park,
  A.~B. Georgescu, J.~Cano, L.~Krusin-Elbaum,
\newblock \emph{Nat. Phys.} \textbf{2020}, \emph{17}, 1 36.

\bibitem{Zhou2005}
Z.~Zhou, Y.-J. Chien, C.~Uher,
\newblock \emph{Appl. Phys. Lett.} \textbf{2005}, \emph{87}, 11 112503.

\bibitem{interconnects}
X.~Zhang, S.-C. Zhang,
\newblock In T.~George, M.~S. Islam, A.~Dutta, editors, \emph{Micro- and
  Nanotechnology Sensors, Systems, and Applications IV}. {SPIE}, \textbf{2012}
  \urlprefix\url{https://doi.org/10.1117/12.920325}.

\bibitem{Yasuda17}
K.~Yasuda, M.~Mogi, R.~Yoshimi, A.~Tsukazaki, K.~S. Takahashi, M.~Kawasaki,
  F.~Kagawa, Y.~Tokura,
\newblock \emph{Science} \textbf{2017}, \emph{358}, 6368 1311.

\bibitem{Mahoney2017}
A.~C. Mahoney, J.~I. Colless, L.~Peeters, S.~J. Pauka, E.~J. Fox, X.~Kou,
  L.~Pan, K.~L. Wang, D.~Goldhaber-Gordon, D.~J. Reilly,
\newblock \emph{Nat. Commun.} \textbf{2017}, \emph{8}, 1 1836.

\bibitem{Xiao2018}
D.~Xiao, J.~Jiang, J.-H. Shin, W.~Wang, F.~Wang, Y.-F. Zhao, C.~Liu, W.~Wu,
  M.~H. Chan, N.~Samarth, C.-Z. Chang,
\newblock \emph{Phys. Rev. Lett.} \textbf{2018}, \emph{120}, 5 056801.

\bibitem{Rienks}
E.~D.~L. Rienks, S.~Wimmer, J.~S{\'{a}}nchez-Barriga, O.~Caha, P.~S. Mandal,
  J.~R{\r{u}}{\v{z}}i{\v{c}}ka, A.~Ney, H.~Steiner, V.~V. Volobuev, H.~Groiss,
  M.~Albu, G.~Kothleitner, J.~Michali{\v{c}}ka, S.~A. Khan, J.~Min{\'{a}}r,
  H.~Ebert, G.~Bauer, F.~Freyse, A.~Varykhalov, O.~Rader, G.~Springholz,
\newblock \emph{Nature} \textbf{2019}, \emph{576}, 7787 423.

\bibitem{HagmannNJP17}
J.~A. Hagmann, X.~Li, S.~Chowdhury, S.-N. Dong, S.~Rouvimov, S.~J.
  Pookpanratana, K.~M. Yu, T.~A. Orlova, T.~B. Bolin, C.~U. Segre, D.~G.
  Seiler, C.~A. Richter, X.~Liu, M.~Dobrowolska, J.~K. Furdyna,
\newblock \emph{New J. Phys.} \textbf{2017}, \emph{19}, 8 085002.

\bibitem{DSLee13}
D.~S. Lee, T.-H. Kim, C.-H. Park, C.-Y. Chung, Y.~S. Lim, W.-S. Seo, H.-H.
  Park,
\newblock \emph{Cryst. Eng. Comm} \textbf{2013}, \emph{15}, 27 5532.

\bibitem{EremeevJAC17}
S.~Eremeev, M.~Otrokov, E.~Chulkov,
\newblock \emph{J. Alloys Compd.} \textbf{2017}, \emph{709} 172.

\bibitem{OtrokovAFMTI18}
M.~M. Otrokov, I.~I. Klimovskikh, H.~Bentmann, D.~Estyunin, A.~Zeugner, Z.~S.
  Aliev, S.~Ga{\ss}, A.~U.~B. Wolter, A.~V. Koroleva, A.~M. Shikin,
  M.~Blanco-Rey, M.~Hoffmann, I.~P. Rusinov, A.~Y. Vyazovskaya, S.~V. Eremeev,
  Y.~M. Koroteev, V.~M. Kuznetsov, F.~Freyse, J.~S{\'{a}}nchez-Barriga, I.~R.
  Amiraslanov, M.~B. Babanly, N.~T. Mamedov, N.~A. Abdullayev, V.~N. Zverev,
  A.~Alfonsov, V.~Kataev, B.~B\"{u}chner, E.~F. Schwier, S.~Kumar, A.~Kimura,
  L.~Petaccia, G.~D. Santo, R.~C. Vidal, S.~Schatz, K.~Ki{\ss}ner,
  M.~\"{U}nzelmann, C.~H. Min, S.~Moser, T.~R.~F. Peixoto, F.~Reinert,
  A.~Ernst, P.~M. Echenique, A.~Isaeva, E.~V. Chulkov,
\newblock \emph{Nature} \textbf{2019}, \emph{576}, 7787 416.

\bibitem{McQueeneyPRM19}
J.-Q. Yan, Q.~Zhang, T.~Heitmann, Z.~Huang, K.~Y. Chen, J.-G. Cheng, W.~Wu,
  D.~Vaknin, B.~C. Sales, R.~J. McQueeney,
\newblock \emph{Phys. Rev. Mat.} \textbf{2019}, \emph{3}, 6 064202.

\bibitem{McQueeneyPRB19}
J.-Q. Yan, S.~Okamoto, M.~A. McGuire, A.~F. May, R.~J. McQueeney, B.~C. Sales,
\newblock \emph{Phys. Rev. B} \textbf{2019}, \emph{100}, 10 104409.

\bibitem{ChenPRM20}
Y.~Chen, Y.-W. Chuang, S.~H. Lee, Y.~Zhu, K.~Honz, Y.~Guan, Y.~Wang, K.~Wang,
  Z.~Mao, J.~Zhu, C.~Heikes, P.~Quarterman, P.~Zajdel, J.~A. Borchers,
  W.~Ratcliff,
\newblock \emph{Phys. Rev. Mat.} \textbf{2020}, \emph{4}, 6 064411.

\bibitem{OtrokovPRL19}
M.~Otrokov, I.~Rusinov, M.~Blanco-Rey, M.~Hoffmann, A.~Vyazovskaya, S.~Eremeev,
  A.~Ernst, P.~Echenique, A.~Arnau, E.~Chulkov,
\newblock \emph{Phys. Rev. Lett.} \textbf{2019}, \emph{122}, 10 107202.

\bibitem{PRX1}
Y.-J. Hao, P.~Liu, Y.~Feng, X.-M. Ma, E.~F. Schwier, M.~Arita, S.~Kumar, C.~Hu,
  R.~Lu, M.~Zeng, Y.~Wang, Z.~Hao, H.-Y. Sun, K.~Zhang, J.~Mei, N.~Ni, L.~Wu,
  K.~Shimada, C.~Chen, Q.~Liu, C.~Liu,
\newblock \emph{Phys. Rev. X} \textbf{2019}, \emph{9}, 4 041038.

\bibitem{PRX2}
H.~Li, S.-Y. Gao, S.-F. Duan, Y.-F. Xu, K.-J. Zhu, S.-J. Tian, J.-C. Gao, 
W.-H.
  Fan, Z.-C. Rao, J.-R. Huang, J.-J. Li, D.-Y. Yan, Z.-T. Liu, W.-L. Liu, 
Y.-B.
  Huang, Y.-L. Li, Y.~Liu, G.-B. Zhang, P.~Zhang, T.~Kondo, S.~Shin, H.-C. Lei,
  Y.-G. Shi, W.-T. Zhang, H.-M. Weng, T.~Qian, H.~Ding,
\newblock \emph{Phys. Rev. X} \textbf{2019}, \emph{9}, 4 041039.

\bibitem{PRX3}
Y.~J. Chen, L.~X. Xu, J.~H. Li, Y.~W. Li, H.~Y. Wang, C.~F. Zhang, H.~Li,
  Y.~Wu, A.~J. Liang, C.~Chen, S.~W. Jung, C.~Cacho, Y.~H. Mao, S.~Liu, M.~X.
  Wang, Y.~F. Guo, Y.~Xu, Z.~K. Liu, L.~X. Yang, Y.~L. Chen,
\newblock \emph{Phys. Rev. X} \textbf{2019}, \emph{9}, 4 041040.

\bibitem{DengScience20}
Y.~Deng, Y.~Yu, M.~Z. Shi, Z.~Guo, Z.~Xu, J.~Wang, X.~H. Chen, Y.~Zhang,
\newblock \emph{Science} \textbf{2020}, \emph{367}, 6480 895.

\bibitem{Vidal2019}
R.~C. Vidal, A.~Zeugner, J.~I. Facio, R.~Ray, M.~H. Haghighi, A.~U. Wolter,
  L.~T.~C. Bohorquez, F.~Caglieris, S.~Moser, T.~Figgemeier, T.~R. Peixoto,
  H.~B. Vasili, M.~Valvidares, S.~Jung, C.~Cacho, A.~Alfonsov, K.~Mehlawat,
  V.~Kataev, C.~Hess, M.~Richter, B.~B\"{u}chner, J.~van~den Brink, M.~Ruck,
  F.~Reinert, H.~Bentmann, A.~Isaeva,
\newblock \emph{Phys. Rev. X} \textbf{2019}, \emph{9}, 4 041065.

\bibitem{HuSciAdv20}
C.~Hu, L.~Ding, K.~N. Gordon, B.~Ghosh, H.-J. Tien, H.~Li, A.~G. Linn, S.-W.
  Lien, C.-Y. Huang, S.~Mackey, J.~Liu, P.~V.~S. Reddy, B.~Singh, A.~Agarwal,
  A.~Bansil, M.~Song, D.~Li, S.-Y. Xu, H.~Lin, H.~Cao, T.-R. Chang, D.~Dessau,
  N.~Ni,
\newblock \emph{Sci. Adv.} \textbf{2020}, \emph{6}, 30 eaba4275.

\bibitem{chen2020}
B.~Chen, F.~Fei, D.~Wang, Z.~Jiang, B.~Zhang, J.~Guo, H.~Xie, Y.~Zhang,
  M.~Naveed, Y.~Du, Z.~Sun, H.~Zhang, D.~Shen, F.~Song,
\newblock \emph{arXiv} \textbf{2020}, \emph{2009.} 00039.

\bibitem{Shi2020}
G.~Shi, M.~Zhang, D.~Yan, H.~Feng, M.~Yang, Y.~Shi, Y.~Li,
\newblock \emph{Chin. Phys. Lett.} \textbf{2020}, \emph{37}, 4 047301.

\bibitem{HuarXiv2020}
C.~Hu, S.~Mackey, N.~Ni,
\newblock \emph{arXiv} \textbf{2020}, \emph{2008.} 09097.

\bibitem{huan2021}
S.~Huan, S.~Zhang, Z.~Jiang, H.~Su, H.~Wang, X.~Zhang, Y.~Yang, Z.~Liu,
  X.~Wang, N.~Yu, Z.~Zou, D.~Shen, J.~Liu, Y.~Guo,
\newblock \emph{arXiv} \textbf{2021}, \emph{2101.} 10149.

\bibitem{Pauly2012}
C.~Pauly, G.~Bihlmayer, M.~Liebmann, M.~Grob, A.~Georgi, D.~Subramaniam, M.~R.
  Scholz, J.~S{\'{a}}nchez-Barriga, A.~Varykhalov, S.~Bl\"{u}gel, O.~Rader,
  M.~Morgenstern,
\newblock \emph{Phys. Rev. B} \textbf{2012}, \emph{86}, 23 235106.

\bibitem{Dyck03}
J.~S. Dyck, P.~Svanda, P.~Lostak, J.~Horak, W.~Chen, C.~Uher,
\newblock \emph{J. Appl. Phys.} \textbf{2003}, \emph{94}, 12 7631.

\bibitem{Choi04}
J.~Choi, S.~Choi, J.~Choi, Y.~Park, H.-M. Park, H.-W. Lee, B.-C. Woo, S.~Cho,
\newblock \emph{phys. stat. sol. (b)} \textbf{2004}, \emph{241}, 7 1541.

\bibitem{liu2021}
Y.~Liu, L.-L. Wang, Q.~Zheng, Z.~Huang, X.~Wang, M.~Chi, Y.~Wu, B.~C.
  Chakoumakos, M.~A. McGuire, B.~C. Sales, W.~Wu, J.~Yan,
\newblock \emph{arXiv} \textbf{2020}, \emph{2007.} 12217.

\bibitem{MurakamiPRB19}
T.~Murakami, Y.~Nambu, T.~Koretsune, G.~Xiangyu, T.~Yamamoto, C.~M. Brown,
  H.~Kageyama,
\newblock \emph{Phys. Rev. B} \textbf{2019}, \emph{100}, 19 195103.

\bibitem{Ge2021}
W.~Ge, P.~M. Sass, J.~Yan, S.~H. Lee, Z.~Mao, W.~Wu,
\newblock \emph{Phys. Rev. B} \textbf{2021}, \emph{103} 134403.

\bibitem{Riberolles2021}
S.~X.~M. Riberolles, Q.~Zhang, E.~Gordon, N.~P. Butch, L.~Ke, J.~Q. Yan, R.~J.
  McQueeney,
\newblock \emph{arXiv} \textbf{2021}, \emph{2103.} 09335.

\bibitem{li2021}
H.~Li, Y.~Li, Y.-K. Lian, W.~Xie, L.~Chen, J.~Zhang, Y.~Wu, S.~Fan,
\newblock \emph{arXiv} \textbf{2021}, \emph{2104.} 00898.

\bibitem{ZhangDPRL19}
D.~Zhang, M.~Shi, T.~Zhu, D.~Xing, H.~Zhang, J.~Wang,
\newblock \emph{Phys. Rev. Lett.} \textbf{2019}, \emph{122}, 20 206401.

\bibitem{ChenBo19}
B.~Chen, F.~Fei, D.~Zhang, B.~Zhang, W.~Liu, S.~Zhang, P.~Wang, B.~Wei,
  Y.~Zhang, Z.~Zuo, J.~Guo, Q.~Liu, Z.~Wang, X.~Wu, J.~Zong, X.~Xie, W.~Chen,
  Z.~Sun, S.~Wang, Y.~Zhang, M.~Zhang, X.~Wang, F.~Song, H.~Zhang, D.~Shen,
  B.~Wang,
\newblock \emph{Nat. Commun.} \textbf{2019}, \emph{10}, 1 4469.

\bibitem{Lei20}
C.~Lei, S.~Chen, A.~H. MacDonald,
\newblock \emph{Proc. Natl. Acad. Sci.} \textbf{2020}, \emph{117}, 44 27224.

\bibitem{LiuY20}
Y.~Liu, L.-L. Wang, Q.~Zheng, Z.~Huang, X.~Wang, M.~Chi, Y.~Wu, B.~C.
  Chakoumakos, M.~A. McGuire, B.~C. Sales, W.~Wu, J.~Yan,
\newblock \emph{arXiv} \textbf{2020}, \emph{2007.} 12217.

\bibitem{Zhou2020}
L.~Zhou, Z.~Tan, D.~Yan, Z.~Fang, Y.~Shi, H.~Weng,
\newblock \emph{Phys. Rev. B} \textbf{2020}, \emph{102}, 8 085114.

\bibitem{Otrokov2DMat}
M.~M. Otrokov, T.~V. Menshchikova, M.~G. Vergniory, I.~P. Rusinov, A.~Y.
  Vyazovskaya, Y.~M. Koroteev, G.~Bihlmayer, A.~Ernst, P.~M. Echenique,
  A.~Arnau, E.~V. Chulkov,
\newblock \emph{2D Mater.} \textbf{2017}, \emph{4}, 2 025082.

\bibitem{RichardsonSR17}
C.~L. Richardson, J.~M. Devine-Stoneman, G.~Divitini, M.~E. Vickers, C.-Z.
  Chang, M.~Amado, J.~S. Moodera, J.~W.~A. Robinson,
\newblock \emph{Sci. Rep.} \textbf{2017}, \emph{7}, 1 12061.

\bibitem{Jiang2012}
Y.~Jiang, Y.~Y. Sun, M.~Chen, Y.~Wang, Z.~Li, C.~Song, K.~He, L.~Wang, X.~Chen,
  Q.-K. Xue, X.~Ma, S.~B. Zhang,
\newblock \emph{Phys. Rev. Lett.} \textbf{2012}, \emph{108}, 6 066809.

\bibitem{Kellner2017}
J.~Kellner, G.~Bihlmayer, V.~L. Deringer, M.~Liebmann, C.~Pauly, A.~Giussani,
  J.~E. Boschker, R.~Calarco, R.~Dronskowski, M.~Morgenstern,
\newblock \emph{Phys. Rev. B} \textbf{2017}, \emph{96}, 24 245408.

\bibitem{XueSTM20}
Y.~Yuan, X.~Wang, H.~Li, J.~Li, Y.~Ji, Z.~Hao, Y.~Wu, K.~He, Y.~Wang, Y.~Xu,
  W.~Duan, W.~Li, Q.-K. Xue,
\newblock \emph{{Nano Lett.}} \textbf{{2020}}, \emph{{20}}, {5} {3271}.

\bibitem{lai2021defectdriven}
Y.~Lai, L.~Ke, J.~Yan, R.~D. McDonald, R.~J. McQueeney,
\newblock \emph{arXiv} \textbf{2021}, \emph{2102.} 05797.

\bibitem{JSB16}
J.~S{\'{a}}nchez-Barriga, A.~Varykhalov, G.~Springholz, H.~Steiner,
  R.~Kirchschlager, G.~Bauer, O.~Caha, E.~Schierle, E.~Weschke, A.~A. \"{U}nal,
  S.~Valencia, M.~Dunst, J.~Braun, H.~Ebert, J.~Min{\'{a}}r, E.~Golias, L.~V.
  Yashina, A.~Ney, V.~Hol{\'{y}}, O.~Rader,
\newblock \emph{Nat. Commun.} \textbf{2016}, \emph{7}, 1 10559.

\bibitem{Henk2012}
J.~Henk, M.~Flieger, I.~V. Maznichenko, I.~Mertig, A.~Ernst, S.~V. Eremeev,
  E.~V. Chulkov,
\newblock \emph{Phys. Rev. Lett.} \textbf{2012}, \emph{109} 076801.

\bibitem{LeeIPNAS2015}
I.~Lee, C.~K. Kim, J.~Lee, S.~J.~L. Billinge, R.~Zhong, J.~A. Schneeloch,
  T.~Liu, T.~Valla, J.~M. Tranquada, G.~Gu, J.~C.~S. Davis,
\newblock \emph{Proc. Natl. Acad. Sci.} \textbf{2015}, \emph{112}, 5 1316.

\bibitem{ChenNJP2015}
C.-C. Chen, M.~L. Teague, L.~He, X.~Kou, M.~Lang, W.~Fan, N.~Woodward, K.-L.
  Wang, N.-C. Yeh,
\newblock \emph{New J. Phys.} \textbf{2015}, \emph{17}, 11 113042.

\bibitem{BeidenkopfNP11}
H.~Beidenkopf, P.~Roushan, J.~Seo, L.~Gorman, I.~Drozdov, Y.~S. Hor, R.~J.
  Cava, A.~Yazdani,
\newblock \emph{Nat. Phys.} \textbf{2011}, \emph{7}, 12 939.

\bibitem{Pauly2015}
C.~Pauly, C.~Saunus, M.~Liebmann, M.~Morgenstern,
\newblock \emph{Phys. Rev. B} \textbf{2015}, \emph{92}, 8 085140.

\bibitem{Morgenstern03}
M.~Morgenstern,
\newblock \emph{Surf. Rev. Lett.} \textbf{2003}, \emph{10}, 06 933.

\bibitem{SessiNC16}
P.~Sessi, R.~R. Biswas, T.~Bathon, O.~Storz, S.~Wilfert, A.~Barla, K.~A. Kokh,
  O.~E. Tereshchenko, K.~Fauth, M.~Bode, A.~V. Balatsky,
\newblock \emph{Nat. Commun.} \textbf{2016}, \emph{7}, 1 12027.

\bibitem{Rosenberg12}
G.~Rosenberg, M.~Franz,
\newblock \emph{Phys. Rev. B} \textbf{2012}, \emph{85}, 19 195119.

\bibitem{Kirkpatrick}
T.~R. Kirkpatrick, D.~Belitz,
\newblock \emph{Phys. Rev. B} \textbf{2015}, \emph{91}, 21 214407.

\bibitem{FuchsPRB14}
D.~Fuchs, M.~Wissinger, J.~Schmalian, C.-L. Huang, R.~Fromknecht, R.~Schneider,
  H.~v.~L\"{o}hneysen,
\newblock \emph{Phys. Rev. B} \textbf{2014}, \emph{89}, 17 174405.

\bibitem{Sales17}
B.~C. Sales, K.~Jin, H.~Bei, J.~Nichols, M.~F. Chisholm, A.~F. May, N.~P.
  Butch, A.~D. Christianson, M.~A. McGuire,
\newblock \emph{npj Quantum Materials} \textbf{2017}, \emph{2}, 1 33.

\bibitem{Mayer14}
M.~Mayer,
\newblock \emph{Nuclear Instrum. Methods B} \textbf{2014}, \emph{332} 176.

\bibitem{Steiner14}
H.~Steiner, V.~Volobuev, O.~Caha, G.~Bauer, G.~Springholz, V.~Hol{\'{y}},
\newblock \emph{J. Appl. Cryst.} \textbf{2014}, \emph{47}, 6 1889.

\bibitem{Gyorffy1973}
B.~L. Gyorffy, M.~J. Stott,
\newblock In D.~J. Fabian, L.~M. Watson, editors, \emph{Proc. of the Int. 
Conf.
  on Band Structure and Spectroscopy of Metals and Alloys}. Academic Press,
  \textbf{1973} 385.

\bibitem{Geilhufe2015}
M.~Geilhufe, S.~Achilles, M.~A. K\"{o}bis, M.~Arnold, I.~Mertig, W.~Hergert,
  A.~Ernst,
\newblock \emph{J. Phys.: Condens. Matter} \textbf{2015}, \emph{27}, 43 435202.

\bibitem{Anisimov1991}
V.~I. Anisimov, J.~Zaanen, O.~K. Andersen,
\newblock \emph{Phys. Rev. B} \textbf{1991}, \emph{44}, 3 943.

\bibitem{Liechtenstein1987}
A.~Liechtenstein, M.~Katsnelson, V.~Antropov, V.~Gubanov,
\newblock \emph{J. Magn. Magn. Mat.} \textbf{1987}, \emph{67}, 1 65.

\bibitem{Soven1967}
P.~Soven,
\newblock \emph{Phys. Rev.} \textbf{1967}, \emph{156}, 3 809.

\bibitem{Gyorffy1972}
B.~L. Gyorffy,
\newblock \emph{Phys. Rev. B} \textbf{1972}, \emph{5}, 6 2382.

\bibitem{Perdew.prl1996}
J.~P. Perdew, K.~Burke, M.~Ernzerhof,
\newblock \emph{Phys. Rev. Lett.} \textbf{1996}, \emph{77}, 18 3865.

\bibitem{Kurz04}
P.~Kurz, F.~F\"orster, L.~Nordstr\"om, G.~Bihlmayer, S.~Bl\"ugel,
\newblock \emph{Phys. Rev. B} \textbf{2004}, \emph{69} 024415.

\bibitem{ref21}
P.~E. Bl\"{o}chl,
\newblock \emph{Phys. Rev. B} \textbf{1994}, \emph{50}, 24 17953.

\bibitem{ref22}
G.~Kresse, J.~Furthm\"{u}ller,
\newblock \emph{Phys. Rev. B} \textbf{1996}, \emph{54}, 16 11169.

\bibitem{ref23}
G.~Kresse, D.~Joubert,
\newblock \emph{Phys. Rev. B} \textbf{1999}, \emph{59}, 3 1758.

\bibitem{Koelling.jpc1977}
D.~D. Koelling, B.~N. Harmon,
\newblock \emph{Journal of Physics C: Solid State Physics} \textbf{1977},
  \emph{10}, 16 3107.

\bibitem{Grimme.jcp2010}
S.~Grimme, J.~Antony, S.~Ehrlich, H.~Krieg,
\newblock \emph{J. Chem. Phys.} \textbf{2010}, \emph{132}, 15 154104.

\bibitem{Grimme.jcc2011}
S.~Grimme, S.~Ehrlich, L.~Goerigk,
\newblock \emph{J. Comput. Chem.} \textbf{2011}, \emph{32}, 7 1456.

\bibitem{Dudarev.prb1998}
S.~L. Dudarev, G.~A. Botton, S.~Y. Savrasov, C.~J. Humphreys, A.~P. Sutton,
\newblock \emph{Phys. Rev. B} \textbf{1998}, \emph{57}, 3 1505.

\end{thebibliography}
 


\phantom{xxxx}
 

\medskip 
\phantom{xxxx}

\end{document}